\documentclass[preprint,12pt]{elsarticle}
\newcommand{\citefig}[1]{(taken from \cite{#1}, with permission)}

\usepackage{amssymb}
\usepackage{lipsum}
\usepackage{url}

\journal{Nuclear Instruments and Methods A}

\begin{document}

\begin{frontmatter}



\title{Design and Performance of a Universal SiPM Readout System for X- and Gamma-Ray Missions}


\author[first,second]{Merlin Kole}
\author[third,second]{Nicolas De Angelis}
\author[4]{Nicolas Produit}
\author[second]{Philipp Azzarello}
\author[second]{Franck Cadoux}
\author[second]{Yannick Favre}
\author[5]{Jochen Greiner}
\author[second]{Johannes Hulsman}
\author[second]{Vishal Kumar}
\author[6]{Sebastian~Kusyk}
\author[4]{Hancheng~Li}
\author[7]{Dominik Rybka}
\author[second]{Jerome Stauffer}
\author[second]{Adrien Stil} 
\author[8]{Jianchao Sun}
\author[6]{Jan~Swakon}
\author[6]{Damian~Wrobel}
\author[second]{Xin Wu}

\affiliation[first]{organization={University of New Hampshire, Space Science Center},
            city={Durham},
            postcode={03824}, 
            state={New Hampshire},
            country={USA}}

\affiliation[third]{organization={INAF-IAPS, },
            addressline={via del Fosso del Cavaliere 100}, 
            city={Rome},
            postcode={I-00133}, 
            country={Italy}}

\affiliation[second]{organization={University of Geneva, DPNC},
            city={Geneva},
            postcode={1205}, 
            state={Geneva},
            country={Switzerland}}

\affiliation[4]{organization={Geneva Observatory, ISDC, University of Geneva},
            addressline={16, Chemin d’Ecogia}, 
            city={Versoix},
            postcode={1290}, 
            state={Geneva},
            country={Switzerland}}

\affiliation[5]{organization={Max-Planck Institute for Extraterrestrial Physics},
            addressline={Giessenbachstr. 1}, 
            city={Garching},
            postcode={85748}, 
            country={Germany}}

\affiliation[6]{organization={Institute of Nuclear Physics Polish Academy of Sciences},
            city={Krakow},
            postcode={PL-31342}, 
            country={Poland}}

\affiliation[7]{organization={National Centre for Nuclear Research},
            addressline={ul. A. Soltana 7}, 
            city={Swierk},
            postcode={05-400 Otwock}, 
            country={Poland}}

\affiliation[8]{organization={Key Laboratory of Particle Astrophysics, Institute of High Energy Physics, Chinese Academy of Sciences},
            city={Beijing},
            postcode={100049}, 
            country={China}}

\begin{abstract}
The advent of both multi-messenger and time-domain astrophysics over the last decade has seen a large interest in the development of small-scale, cheap, and robust gamma-ray detectors. This has been further encouraged by the availability of CubeSat platforms. Of particular interest are detectors capable of producing spectral and localization measurements of X and gamma-ray transients to allow for accurate follow-up measurements at different wavelengths. A vast number of the instruments developed for such purposes in the last years use a combination of scintillators and Silicon Photomultipliers (SiPMs) for photon detection. Here, we present the design, performance, and space qualification of a readout system capable of reading out 64 SiPM channels. This low-power and low-cost system was originally designed for the POLAR-2 mission, a large scale gamma-ray polarimeter. However, its flexible design makes it equally suitable for use on various CubeSat missions. The system was found to perform well when reading out both plastic and high Z scintillators using a total of 1.8~W. The space qualified design furthermore relies on commercial off-the-shelf components, thereby also removing most international export issues. In this paper, we will present the overall design, the performance of the electronics, its performance when reading out various scintillators and the successful space-qualification of this design.
\end{abstract}




\end{frontmatter}




\section{Introduction}
\label{introduction}

The joint Gravitational Wave (GW) and Gamma-ray Burst (GRB) detection of GW170817/GRB170817A indicated that gamma-ray detectors are vital in the field of multi-messenger astrophysics \cite{170817}. These detectors need to be capable of detecting weak transients and alerting the community with temporal, spectral and localization information. This increase in interest in developing such detectors coincided with the emergence of SiPM technology. Traditional gamma-ray detectors, such as BATSE \citep{Fishman1994ApJS...92..229F}, Fermi-GBM \citep{Meegan2009ApJ...702..791M}, and POLAR \citep{PRODUIT2018259}, were based on scintillators read out by Photo-Multiplier Tubes (PMT). Downsides of PMTs are that they are bulky and mechanically fragile. In addition, such instruments required complex high-voltage power supplies capable of delivering high voltage values typically around kV. The development of SiPMs mitigates all such issues as they are compact, mechanically robust, and require typical operating voltages below 100~V. These properties make SiPMs ideal for replacing PMTs, especially for small-scale space-based detectors. As such, a combination of SiPMs and scintillators have been used or proposed for missions such as:
GRBAlpha \citep{GRBAlpha}, VZLUSAT-2 \citep{VZLUSAT-2}, BurstCube \citep{BCube}, Glowbug \citep{Glowbug} and GALI \citep{GALI} which use SiPMs with CsI, COMCUBE \citep{COMCUBE} and EIRSAT-1 \citep{Eir} which are designed to use SiPMs with CeBr$_3$,  GARI \citep{GARI}, GRID \citep{GRID} and GTM \citep{GTM} which use SiPMs coupled to GAGG, MAMBO \citep{MAMBO} which uses it together with BGO and SIRI \citep{SIRI} which couples it to SrI$_2$. Apart from these relatively small-scale instruments, larger-scale missions like GECAM \citep{Lv2018JInst..13P8014L}, StarBurst \citep{Starburst} and POLAR-2 \citep{POLAR-2_paper} are also using SiPMs in their design.

In the case of POLAR-2, a total of 6400 scintillators are coupled to an equal number of SiPM channels. This large number of channels required the development of a low-power electronics design. In addition, to keep such a large mission within a manageable budget, as well as to avoid typical export issues that come with high-end space-qualified components, it was decided to only use commercial off-the-shelf (COTS) components. The final design of this front-end readout electronics (FEE), which will be presented in detail in the following section, allows for the readout of 64 SiPM channels. The typical power consumption of the system is 1.8~W, while the cost, excluding SiPMs, is approximately 3000~USD. One exception to the COTS components is the ASIC used. It should, however, be noted that, at least for POLAR-2, this component is free from export restrictions and commercially available.

While the FEE was designed with the primary goal in mind to be used for POLAR-2, it was soon recognized that it could be used for a wider variety of SiPM-based detectors. For this purpose, the design was kept flexible. For example, while in POLAR-2 it is used to read out a tightly packed plastic scintillator array, it was shown to also function well when reading out different types of scintillators such as GAGG, or CeBr$_3$. This indicates that, with simple modifications, the layout of the scintillar array and the SiPM types attached can be changed relatively easily with respect to the POLAR-2 one to, for example, to produce a design optimized for spectrometry. The main components of the FEE, as well as the overall design, were furthermore space-qualified during a range of campaigns. As such, the design is currently being proposed to be used on a secondary payload of POLAR-2, the Broad energy-band Spectrum Detector (BSD), an instrument to measure the cosmic X-ray background CXBe \citep{CXBe_paper, CXBe_ICRC} and it is being considered for use on the future eXTP mission \citep{eXTP} as a Wide-field Wide energy-band Camera (W2C). Finally, studies to investigate whether the design is suitable for use with Linearly Graded SiPMs from FBK (LG-SiPMs) \citep{LG-SiPM} are currently ongoing by the authors.

In this manuscript, we will present the overall design of the FEE in section \ref{sec:design}. This is followed by the details on the performance of the electronics of the first prototypes in section \ref{sec:elec_perf}. The system was extensively tested to read out Hamamatsu S13361 MPPCs/SiPMs coupled to plastic scintillators, the performance of which is summarized in section \ref{sec:plastic_perf}. The performance after replacing the plastic with GAGG is described in section \ref{sec:GAGG_perf}. Finally, the various space qualification tests and their outcomes are presented in section \ref{sec:space_qual}. The manuscript finishes with a discussion on potential future applications of the design and shortcomings of the current design.

\section{Overall Design}\label{sec:design}

\subsection{Design constraints}

When designed for POLAR-2, the readout system was required to be capable of reading out 64 SiPM channels at low power, low cost and with minimal heating of the SiPMs by the electronics. The low power constraint here stemmed from the power limit of POLAR-2 being 300~W.  To optimize the sensitivity of the polarimeter this power needed to be used for as many SiPM channels as possible, the ambitious aim being a total of 6400 channels. This would translate to 100 read-out boards to be used. When taking into account the power required for the back-end electronics, voltage conversion efficiencies and heating, this leaves approximately $\approx$ 2~W for each read out board. The requirement for minimal cost was, as can be expected, related to a tight budget.  Finally, the heating requirements stems from SiPMs performing better at lower temperatures. For POLAR-2 a target temperature of $0\,\mathrm{C^o}$ was set as it provides an acceptable noise level while keeping the SiPM well within the designed operation range. In order to allow for further lowering the temperature of the SiPMs by several degrees and actively stabilizing the temperature of the SiPMs, the board is capable of driving a Peltier element which can be placed on the back of the SiPMs. In addition, it can also drive power resistors which have the goal of heating the SiPMs. This capability was added to allow for thermal annealing of radiation damage \citep{DeAngelis_2023}. The heating power reserved for both the cooling and heating in POLAR-2 is approximately 1~W. Moreover, it should be noted that, while for POLAR-2 the Peltier element is used for cooling only\footnote{A design decision based on limited space on the PCB.}, one could redesign its control logic and power supply such that it can also heat the SiPMs, thereby removing the need for having the power resistors. 

A final consideration in the design, especially important for space missions, is the mass of the design. For POLAR-2 the total allowed mass of the payload is $300\,\mathrm{kg}$ most of which is taken up by the overall mechanics and the mass of the scintillators. The front-end boards discussed here and their mechanics weigh only 170 grams a piece and therefore only make up $17\,\mathrm{kg}$ of the full payload. When used for a Cubesat like mission, this mass is of course a more important issue, where it could be reduced, for example by further minimizing the  aluminium mechanics and the copper cooling tube.

\begin{figure}[ht]
\centering
 \includegraphics[width=16 cm]{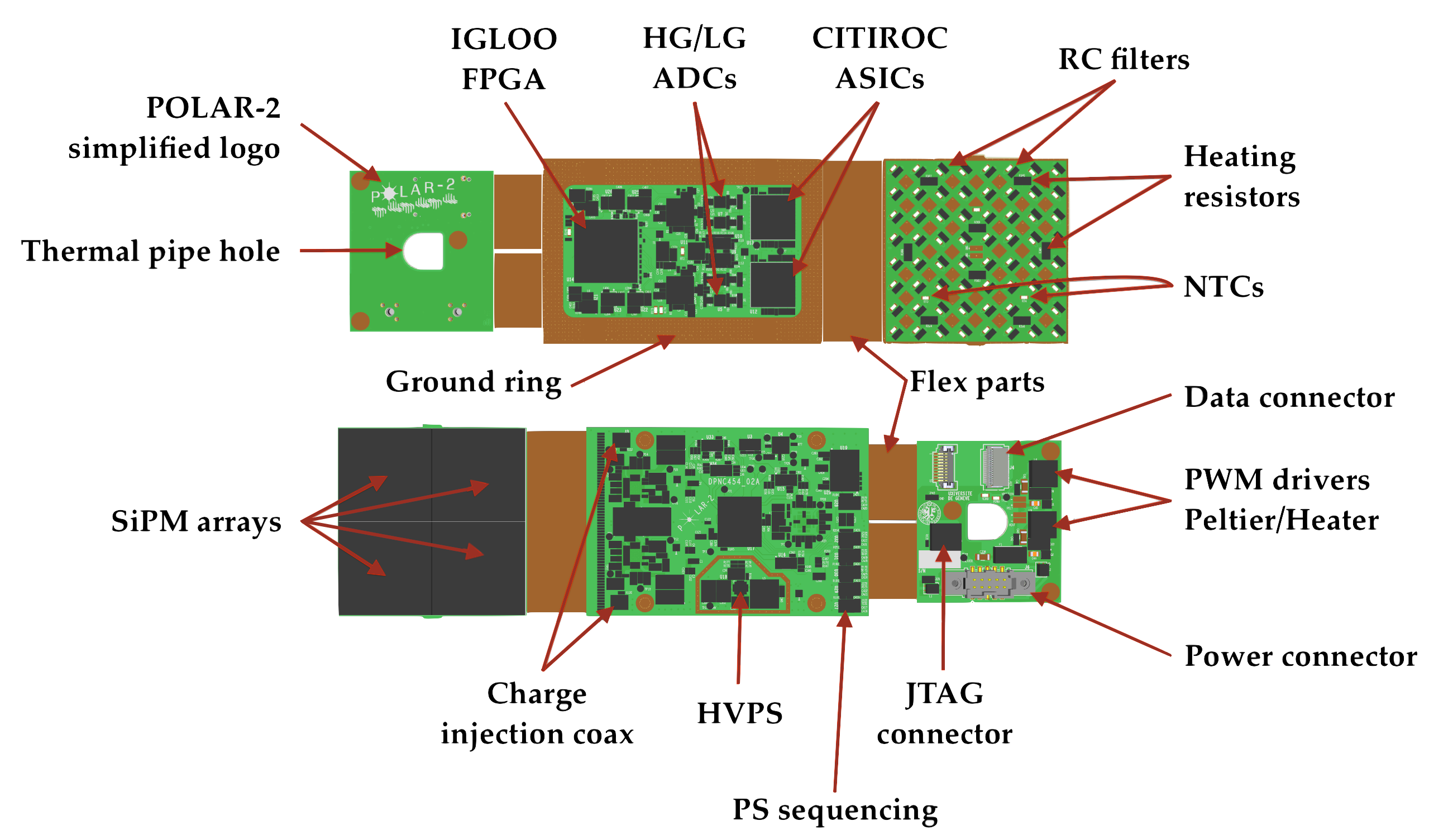}
 \caption{The FEE design in its current form as used for POLAR-2. The two sides are shown here (top and bottom). The leftmost PCB contains the SiPMs, filters for the SiPM bias and heating resistors. This board is connected through a flex to the main PCB which contains the majority of the components. The right PCB contains the connectors as well as drivers which can be used to power a potential Peltier element.  \citefig{NDA_thesis}}
 \label{fig:FEE_design}
\end{figure}

\begin{figure}[h!]
\centering
 \includegraphics[width=12 cm]{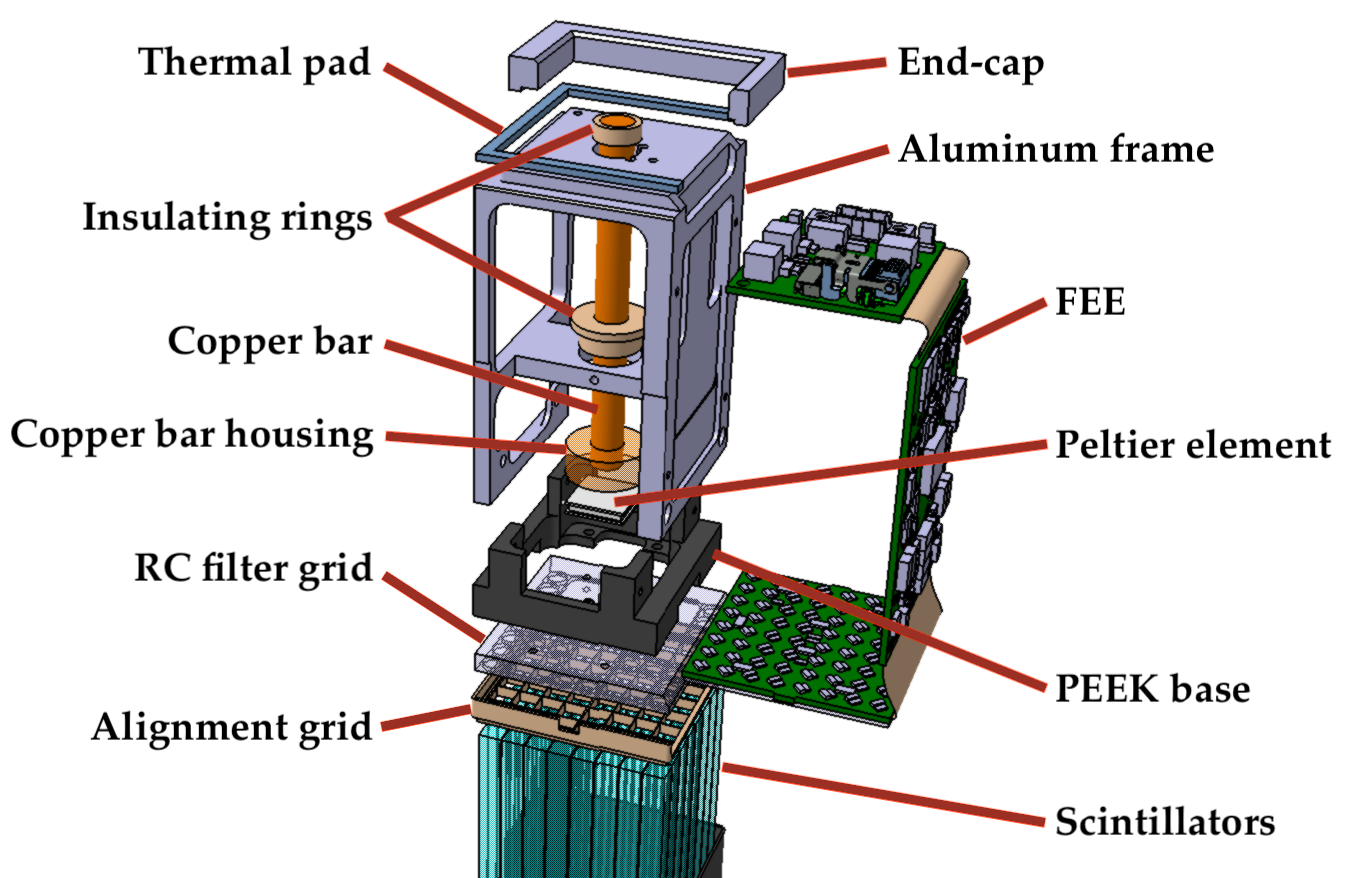}
 \caption{Image of how the front-end electronics is, as of the design in 2023, mechanically integrated in the POLAR-2 detector. The SiPM PCB and connector PCB parts are placed with a 90$^\circ$ angle with respect to the rigid PCB part which holds the main components. This main PCB is mechanically coupled to an aluminum frame which forms the start of a thermal path to a radiator. In order to thermally insulate the SiPM PCB from the aluminium the two are separated by a PEEK base. On the other side the SiPMs are connected to a plastic grid which provides precise alignment to the scintillator array (shown in blue at the bottom). In this design the SiPM PCB is furthermore coupled to a Peltier element connected to a copper bar which forms the thermal path to the aluminium.}
 \label{fig:thermomechal_design}
\end{figure}

To reduce heating of the SiPMs by the electronics, it was decided to place these electronics on a PCB separate from the SiPMs, thereby allowing to thermally isolate both parts. The two PCBs are connected through a flexible PCB. Finally, for mechanical convenience, it was then decided to place the various connectors on a final third PCB which again is connected through a flex PCB. The full design can be observed in figure \ref{fig:FEE_design}, while the way it is mechanically mounted in POLAR-2 is shown in figure \ref{fig:thermomechal_design}. 

In order to minimize the power consumption and simplify the design, it was decided to read out the SiPMs using well-tested commercially available ASICs. For this purpose, the Citiroc 1A ASIC by Weeroc \citep{Citiroc} was chosen. These 32-channel ASICs have two separate amplifiers for each channel, as well as two separately adjustable thresholds. This offers a large flexibility when designing the trigger logic. More details on the usage of these ASICs is provided in the next section.

As for POLAR-2 a total of 100 of these front-ends are required a large level of autonomy and scalability is desired. For this purpose, each FEE has its own FPGA which handles the trigger logic and control of the various components on the board. As such, the system can be used on its own using a simple back-end or, when using a more complex back-end design, tens to hundreds can be used within a single detector. To accommodate for the latter scenario, the system has a total of 8 LVDS pairs, which can be used to provide a flexible communication while keeping the requirements for the back-end relatively simple. In the current design, these LVDS pairs are used as follows:

\begin{itemize}\setlength{\itemsep}{-0.3ex plus0.1ex}
    \item Data Write: Used for writing data from the back-end to the front-end.
    \item Data Read: Used for reading data to the back-end from the front-end.
    \item Sync: Used to provide a clock signal to the front-end.
    \item Trig-Out: Used for the system to signal it has a valid event for triggering.
    \item Trig-In: Can be used to force the system to trigger by the back-end.
    \item Spare 1 out: Can be used to, for example, signal a large energy deposition likely to come from a cosmic-ray interaction.
    \item Spare 2 out: Can be used to, for example, signaling an event with high multiplicity.
    \item Spare 1 in: Can be used to signal something to the front-end.
\end{itemize}

Finally, for simplicity, the system is setup to operate on a single voltage which is 3.8~V. The board provides the high voltage at tens of volts (see \ref{sec:HVPS}) for the SiPMs by itself to further reduce dependency on a back-end.

\subsection{General electronics design}

\subsubsection{PCB Design}

To accommodate the thermal constraints described above, as well as for mechanical purposes, it was decided to design the system as a rigid flex PCB. The design contains 3 rigid parts connected by 2 flex parts. The flex parts have a thickness of 0.16 mm and a length of 12 mm. This length allows the 3 rigid boards to be placed at 90 degree angles with respect to one another. The rigid parts are 10 layer PCBs with a thickness of 1.6 mm. The outer two layers are used for components, 3 layers are used for ground, 2 for LVDS, 1 for digital power and 2 for the analog part (which is additionally on the outer layer).

\subsubsection{SiPM PCB}

One of the rigid parts contains the SiPMs on one side, while the other side is occupied by filters for the bias voltage, as well as temperature sensors (NTC) and heating resistors. The latter are placed such that the SiPMs can be heated temporarily to speed up the thermal annealing of radiation damage \citep{DeAngelis_2023}. 

In the baseline design, this PCB provides space for 4 SiPM arrays of 4$\times$4 S13361-6075 MPPCs. For POLAR-2 this SiPM type was selected as the detector requires a densely packed readout of plastic scintillator bars. The use of arrays allows minimizing the pitch between scintillators, while the use of the 75 $\mu m$ microcell SiPMs provides the largest possible photo-detection efficiency, typically required for plastic scintillators. As will be discussed in section \ref{sec:GAGG_perf}, smaller size microcell SiPMs are recommended when using more luminous scintillator types. This would require no change on the design. If however, the size of the SiPMs (here $6\times6\,\mathrm{mm^2}$), or their pitch needs to be changed, a redesign of the PCB is required. However, as this PCB is basically independent of the significantly more complex part which contains the main electronics, this is relatively straight-forward. Only the dimensions need to be modified while the interface with the main PCB can remain the same.

\subsubsection{Connector PCB}

The system contains an additional PCB which mainly hosts the various connectors. In the POLAR-2 design, this part additionally hosts drivers for both a Peltier element and heating resistors which can be used to alter the temperature of the SiPMs. For POLAR-2 current thermal simulations indicate that the system can raise and lower the temperature by aout $5\,\mathrm{C^o}$. These drivers are operated directly from the FPGA hosted on the main PCB. For the usage of the Peltier and the heater an additional power supply is required. In the baseline design this is 1V which is the only power source not created directly from the 3.8~V input which powers all the other components on the board. Providing the external 1V power is optional here as it is only used for powering the potential Peltier or heating elements. 

Both the 3.8 and the 1V input power are provided to the power connector, which is of the type S2SD 2.00 mm Tiger Eye\texttrademark{} from Samtec \footnote{\url{https://www.samtec.com/products/s2sd}}. This connector is mechanically robust, while also allowing to couple to power cables from Samtec which do not contain any materials prone to outgassing. It is worth noting that in an earlier prototype, the FTS-105-01-L-DV connector from Samtec was used. Although this performed well during all space qualification tests, it coupled to cables which contained significant amounts of PVC which caused concerns for thermal vacuum tests. 

The data connection, the different pairs of which have been described in the previous section, is provided through a ECUE FireFly\texttrademark{} from Samtec. This was selected based on the robust design required to survive vibrations and shock, while also being low cost and easy to use.

Finally, this PCB contains a JTAG (Joint Test Action Group) connector, which is used to program the FPGA.

\subsubsection{FPGA}

The FPGA was selected based on heritage from the POLAR mission \citep{PRODUIT2018259}. The POLAR mission used 25 IGLOO FPGAs from Microsemi, all of which survived 2.5 years in low Earth orbit without any issues\footnote{While POLAR stopped operation after 7 months due to a high voltage issue, the electronics still functioned without issues until the instrument was de-orbited in summer of 2019.}. Based on this performance, for POLAR-2 a slightly more powerful version, the IGLOO AGL1000V2-FGG256 from Microchip Technology was selected. Apart from the heritage of IGLOO FPGAs from POLAR, this FPGA was selected based on the low power consumption, the large number of I/O's (294) and its low cost (order of $100\,\mathrm{USD}$). The large number of I/O's is required to deal with the significant number of signals from the two ASICs. Overall, only one pair of I/O's is not occupied in the baseline design.

Using $35\%$ of the logic gates and $72\%$ of the available memory the FPGA currently performs the following tasks:

\begin{itemize}\setlength{\itemsep}{-0.3ex plus0.1ex}
    \item Control of the High Voltage Power Supply
    \item Temperature correction for bias voltage
    \item Readout and control of the house-keeping ADC
    \item Readout and control of signal ADCs
    \item Control of baseline DACs
    \item Slow control of the ASICs
    \item Handling the trigger logic (an example will be provided in section \ref{sec:ASIC}).
    \item Science Data Packaging
    \item House-keeping Data Packaging
    \item Communication with the back-end electronics
    \item Data Compression
    \item Control of the Peltier/Heater Drivers
\end{itemize}

The bias voltage correction for the SiPMs is performed based on a temperature readout on the SiPM board. The FPGA can use a lookup table containing the drift in breakdown vs. temperature converted to the DAC value provided to the high voltage power supply (described in section \ref{sec:HVPS}). As such, the FPGA can adjust the provided bias voltage to the SiPMs for drifts in temperature once per second.

The typical data packages handled by the FPGA contain 128 ADC values (both high gain and low gain ADC from the ASIC), a time stamp, the trigger patterns (both for the time and charge threshold). The time stamp here can be based on an internal clock, or from an external clock provided over the sync line. To reduce the data packet size data compression can be performed. This consists of suppressing all ADC values which are below an adjustable threshold. 

\subsubsection{ASICs}\label{sec:ASIC}

The ASIC chosen for this readout system is the Citiroc 1A by Weeroc \citep{Citiroc}. This ASIC was designed to readout a total of 32 SiPMs. It was chosen for this project based on the large flexibility it allows for setting up trigger logics, its capabilities to fine-tune several parameters for each of the 32 detector channels and its Technology Readiness Level of 8.

The ASIC has 2 adjustable amplifiers for each of the 32 channels producing a high gain and a low gain output. Latching of the signal is performed based on comparisons with the various thresholds, or, can be forced externally. The high gain output can be used to measure the energy range with individual photo-electrons, while the low gain can be used to extend the dynamic range into the hundreds of keV or even MeV energy range. Both the high gain and low gain signals are provided in multiplexed outputs. The input signal can be tested against two individual adjustable threshold levels, one called the 'charge threshold', the other called the 'time threshold'. Although here these two threshold values are used simply as two different levels against which we compare the input signal, and not specifically for charge or time measurements, we will continue to follow this naming convention used by the ASIC manufacturer. If one of the channels has exceeded the charge threshold this information is provided on one output line. In case the time threshold is exceeded by a channel the full trigger pattern is provided in multiplexed form. Details on the complex trigger scheme implemented for the POLAR-2 mission are provided in section \ref{sec:trig_scheme}.

Apart from its capability to provide a complex, flexible trigger logic, the Citiroc 1A also allows to fine-tune a range of parameters for the 32 channels. These parameters, and whether they are applied to all 32 channels (common), or to individual channels are described in table \ref{tab:parameters}.

\begin{table}[!h]
\begin{center}
\caption{Your caption.}
\label{tab:parameters}
\begin{tabular}{||c | c|  c ||} 
 \hline
 Parameter & Channel/Common & Bits  \\ [0.5ex] 
 \hline\hline
 High Gain Amplification & Channel & 8 \\ 
 \hline
 Low Gain Amplification & Channel & 8 \\ 
 \hline
 Charge Threshold & Common & 10 \\
 \hline
 Charge Threshold correction & Channel & 6  \\
 \hline
 Time Threshold & Common & 10 \\
 \hline
 Time Threshold Correction & Channel & 6  \\
 \hline
 Bias Voltage Correction & Channel & 10 \\ [1ex] 
 \hline
\end{tabular}
\end{center}
\end{table}

The bias voltage correction allows for a common bias voltage can be applied to the 32 channels. The corrections are then used to accommodate for variations in the breakdown between various SiPMs. As such variations, within a single type, are typically of the order of hundreds of mV, the 2.5~V option is normally sufficient to produce a uniform over-voltage. The 4.5~V option can be of use in case, for example, one wants to operate different channels at different over-voltage levels.

It should finally be noted that since the start of the development of this front-end, the Radioroc 2 ASIC has been developed by Weeroc \citep{Radioroc}. This ASIC has all the features of the Citiroc 1A while allowing it to read out 64 channels compared to the 32 from the Citiroc. An upgrade to this ASIC in future versions is therefore being considered.

\subsubsection{Current Trigger Scheme}\label{sec:trig_scheme}

The two different amplifiers on the ASIC as well as the two thresholds allow to setup a complex trigger scheme. The system was originally designed for a gamma-ray polarimeter for which coincident events between at least 2 channels, out of the total of 6400, are of interest. These 2 detector channels can be within one 64-channel front-end readout, or can be between 2 different front-ends. To accommodate this, the time threshold and charge threshold are used in parallel to identify different trigger types. In addition, for this experiment, the system is setup to allow to force a trigger externally through an LVDS line which is controlled by a back-end electronics. In all cases, the FPGA forces both ASICs to enter peak sensing mode for 90 ns. During this period each of the 32 channels will record the highest peak, latch on this, and pass this on to the multiplexed outputs (both for HG and LG). The output is passed to ADCs, described later in this section, which convert these pulse heights into digital values before passing these to the FGPA.

External triggers can also be provided to the FPGA over an LVDS line. These are used to readout the pedestal or baseline levels of all 64 channels. These events allow to calibrate both the baseline as well as the noise levels as will be discussed later on. For science data, the charge trigger is setup at a higher level such that, in case of 1 channel on the front-end having a significant energy deposition, the FPGA forces peak sensing to commence on all channels and initiates a readout of the ADCs. In addition, the FPGA sends out a signal on the 'trig out' LVDS line to the back-end. This starts a clock on the back-end  during which it waits for data from any other front-end which it combines into overall event packets for the full instrument. To avoid chance coincidence events, for POLAR-2, the clock is set to 100~ns. This allows to detect coincident events within the full detector, while, by using the higher of the two threshold, avoids dark noise induced dead time.

The time threshold is instead setup to record coincident events within a single front-end. The time threshold is set at a level within the dark noise region (typically around 4 or 5 photo-electrons). However, to avoid significant dead time, the ADC readout is only initialized when at least 2 of the time trigger lines are high. This check is performed by the FPGA using the multiplexed time threshold outputs from the 2 Citirocs. This check is performed within 90 nanoseconds. The multiplicity counting by the FPGA also allows to reduce charged particle induced events, as these typically produce events within the detector with a high number of detector channels triggering. In case that within a single front-end the multiplicity is high, the FGPA uses one of the spare LVDS trigger lines to indicate this to the back-end, thereby allowing it to veto any triggers provided by other front-ends.

\begin{figure}[h!]
\centering
 \includegraphics[width=10 cm]{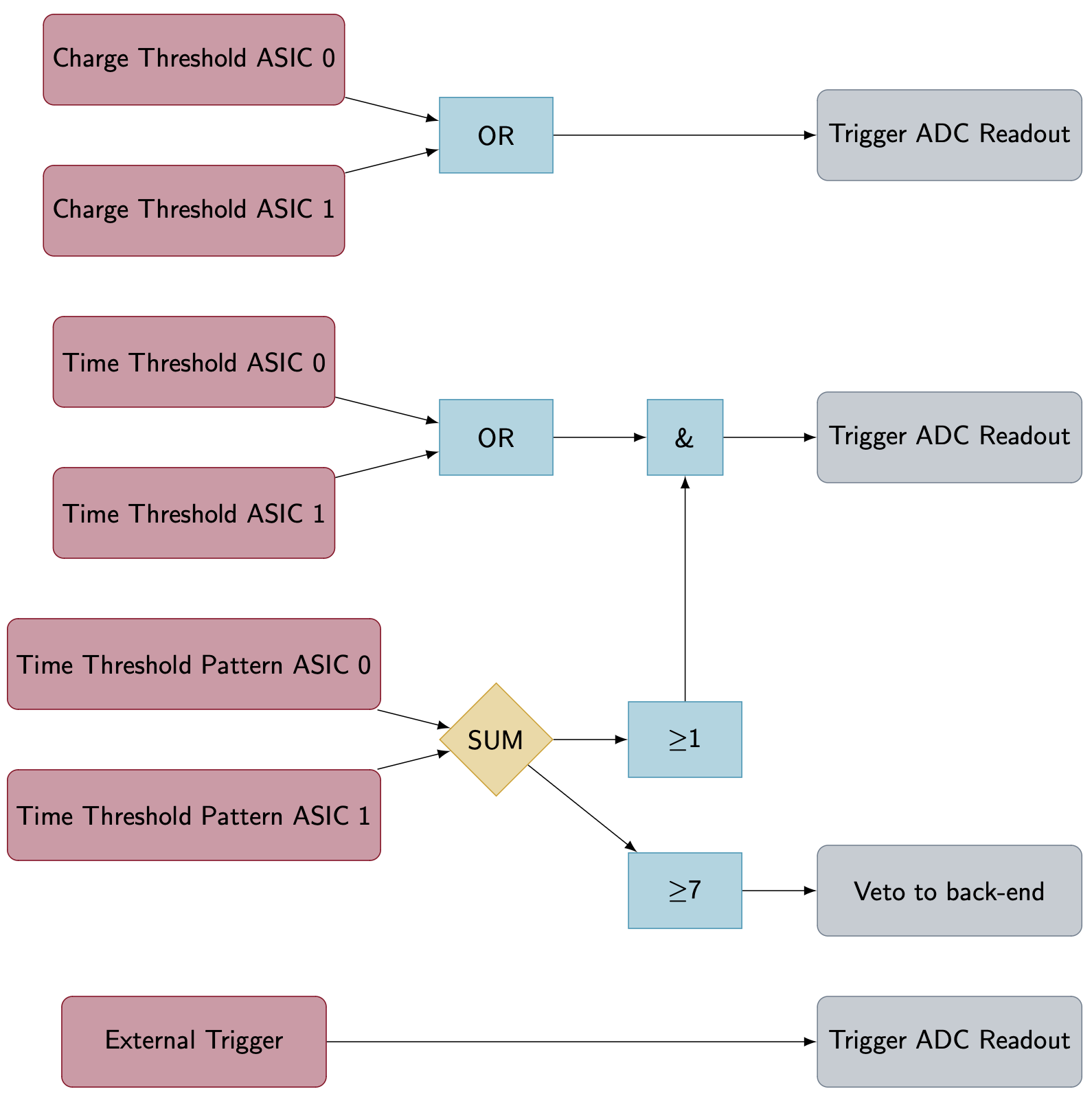}
 \caption{A schematic overview of the trigger logic implemented using the 2 Citiroc ASICs for POLAR-2. The charge threshold is set above the dark noise level, while the time threshold is set lower, typically at the 4 or 5 photo-electron level.}
 \label{fig:trig_logic}
\end{figure}

An overview of the POLAR-2 trigger logic in a detector module is shown in figure \ref{fig:trig_logic}. This logic can easily be adapted for different purposes. For example, when reading out individual channels in a spectrometer one might only be interested in the charge threshold-induced triggers, thereby allowing to remove the time threshold part. Alternatively, in case the 64 channels are used to read out a single scintillator crystal, one might only want to see high multiplicity events, which can be achieved by slightly modifying the time threshold part. Since, as mentioned above, the FPGA can handle additional tasks the trigger logic could be made more complex. In addition, the FPGA could be used to also handle tasks such as auto-calibration or histogramming.

For any of the above described trigger types, the ADC readout is initiated. As the Citiroc 1A does not have its own ADC, each Citiroc is connected to the LTC2263IUJ-12 PBF 2-channel ADC. This ADC allows for simultaneous sampling of 2 14-bit ADC channels, thereby allowing to convert both the high gain and the low gain for the 32 channels. The readout of the two different gain lines for the 32 channels takes approximately 9~$\mu s$ as is shown later in section \ref{sec:dead}. To adapt the baseline or pedestal levels of the ASIC readout, two DACs are used. These DACs are operated by the FPGA and can be used to, for example, equalize the baselines of the two different ASICs.

The trigger logic presented here is optimized to readout a detector with a large number of readout channels in case at least two of these channels exceed a threshold. When used as a spectrometer, this design can be simplified by, for example, only using one of the two thresholds. In addition, it can also be made more complex by, for example requiring at least one channel to be above the charge threshold with a second being required to be above the time threshold. Such a solution can be useful when different types of scintillators are connected to one front-end.

\subsubsection{HVPS}\label{sec:HVPS}

The bias voltage for the 64 SiPMs is provided by the LT3482 DC/DC converter from Linear Technology. This DC/DC can provide a stable voltage between 0 and 90 V, thereby allowing to accommodate all commonly used SiPM types (where the breakdown is typically either around the 30 to 40 V region or the 60 V region). The typical current drawn for one channel of the S13361-6075 SiPM type at standard operating voltages is of the order of several $\mu A$, thereby resulting in a current of hundreds of $\mu A$ for the 64 channels. 

It should be noted that after radiation damage the dark current will significantly increase, see e.g.  \cite{MITCHELL2022167163, DeAngelis_2023, Merzis24154990}. For POLAR-2, the yearly radiation dose has been estimated to be 0.0789 Gy/yr (at an altitude of 383 km) through simulations. This would result in an increase in dark current of a factor of 7.2 (for the 6025 or 6050 types would be a factor 15.7 and 15.4, respectively \citep{DeAngelis_2023}). After one year, the current in one readout system would therefore increase to the order of $1-10 mA$, while after several years, when correcting for annealing, this could increase to the $100\,\mathrm{mA}$ level. The LT3482 DC/DC converter can provide up to 200 mA at 70 V over voltage. Therefore, it allows to continue to power the SiPM array even after significant levels of radiation damage.

\subsubsection{House Keeping}

The increase in current levels due to radiation damage also requires a compromise for the current levels which one wants to read out the housekeeping data. The current to the SiPMs is one of the various values monitored during data taking. This housekeeping data is collected by a separate, slow 12-bit ADC which is controlled by the FPGA. The housekeeping data comprises 4 temperatures: one readout near the FPGA, while the other 3 are distributed over the SiPM board. The need for 3 readouts is to cover potential temperature gradients over the board which can result in unwanted variations over-voltages. In addition, the bias voltage provided to the SiPMs is provided along with the current to the SiPMs. As these currents can vary over several orders of magnitude, a single ADC does not suffice to cover the full range. In the baseline design, the system is set up to read currents in the hundreds of $\mu A$ to the tens of mA. Therefore, the system is able to monitor increases due to radiation damage and is able to potentially measure IV curves after radiation damage. Before radiation damage, when the currents are below the sensitivity range, one relies on the IV curves measured before installation of the SiPMs.

\section{Electronics Performance}\label{sec:elec_perf}

\subsection{Power Consumption}\label{sec:Power}

The system operates at 3.8~V. With the full system powered, excluding a potential Peltier element, the total current drawn is 455~mA which translates to 1.7~W. This, based on calculations performed and tested during the design phase, can be broken down as shown in Table \ref{tab:power}. In addition, as the board only gets 3.8~V as input and has to supply different voltages to the various components, efficiencies between $73\%$ (for the 1.8~V) and $92\%$ (for the 2.5~V) apply. The final result is a power consumption of 1.78~W. As discussed in section \ref{sec:noise}, the solution used to power the various components does not introduce a significant amount of noise into the signal lines.

\begin{table}[h!]
\begin{center}
\label{tab:power}
\begin{tabular}{||c || c | c | c||} 
 \hline
 Component & Voltage (V) & Current (mA) & Power mW \\ [0.5ex] 
 \hline\hline
 2 ASICs & 3.3 & 130 & 430 \\ 
 \hline
 Signal ADCs & 1.8 & 162 & 292 \\
 \hline
 HK ADCs & 3.3 & 17 & 55 \\
 \hline
 DAC & 3.3 & 2 & 7 \\
 \hline
 FPGA core & 1.5 & 40 & 60 \\
 \hline
 FPGA LVDS & 2.5 & 80 & 200\\ 
 \hline
 LVDS TX & 3.3 & 23 & 76\\
 \hline
 LVDS RX & 3.3 & 25 & 83\\
 \hline
 Analog comp & 3.3 & 80 & 264\\
 \hline
 SiPM HVPS & 3.8 & 59 & 225\\
 \hline
  \hline
 Total & 3.8 & 445 & 1696 \\ [1ex] 
 \hline
 \hline
\end{tabular}
\end{center}
\caption{A breakdown of the power consumption of the components on the board.}
\end{table}

\subsection{Gain Variations}\label{sec:noise}

In this design, the signals from the SiPMs arrive at the ASICs where they are amplified. This is done by both a high gain and a low gain amplifier, therefore two amplified signals are produced with different amplitudes. These are finally processed in ADCs which produce the digital values processed in the FPGA. In an ideal detector, one records similar ADC values when two equal input charges are injected in different detector channels. Apart from an equal response of the different channels in the electronics, this also requires an equal sensitivity of the scintillators and SiPMs. While the former can be achieved through careful quality control of the scintillator production and placement, the latter can generally be achieved by optimizing the voltages applied to the various SiPM channels to equalize variations in their gain. For the electronics discussed here, it is important to understand whether an equal charge from the SiPM results in a similar ADC value. For this purpose, charges produced using a pulse generator were injected directly into the electronics just before the ASICs, thereby bypassing the SiPMs and allowing to study the electronics independently. For this test, charges ranging from 20~mV to 360~mV, with incrementing steps of 20~mV were injected in all channels of one ASIC. 

\begin{figure}[h!]
\centering
 \includegraphics[width=15 cm]{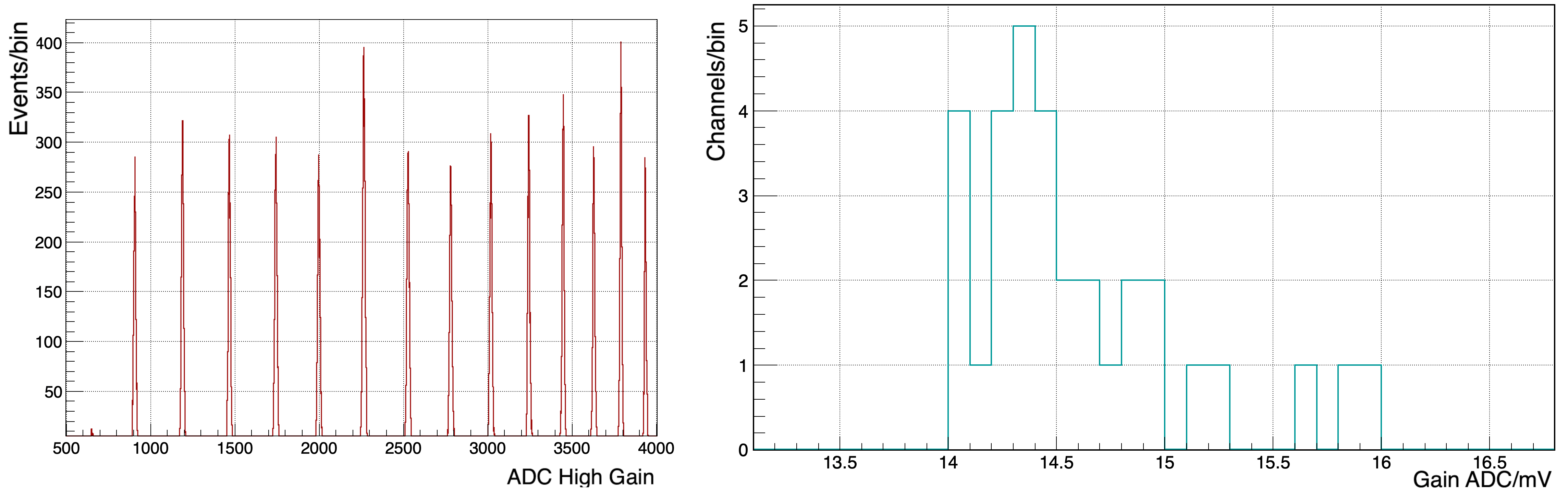}
 \caption{\textbf{Left:} The high gain ADC spectrum from channel 0 for a measurement where charges were injected directly into the ASIC. For this test, charges ranging from 20~mV to 360~mV, with incrementing steps of 20~mV were injected. \textbf{Right:} The gain in ADC/mV for the 32 channels of the tested ASIC as measured using direct charge injection.}
 \label{fig:charge_inj}
\end{figure}

The resulting spectrum, containing all the injected charges of the scan, for one channel are shown in the left panel of figure \ref{fig:charge_inj}. In this figure we can see that each injected charge results in a gaussian distribution of ADC values. The width of this is an indication of the electronic noise which will be discussed in detail in section \ref{sec:noise}. In addition, we see that the lowest charge injection is seen at around 900 ADC, this is a result of a baseline, or pedestal (around 650 ADC for this channel). Finally, it can be observed that at high ADC values the charge induced peaks are closer to each other than at low ADC values. This is an effect of the saturation of the amplifier, an effect which will be discussed and quantified in detail in section \ref{sec:non_linear}. 

The distance in ADC between the peaks in the linear regime (below ~2000 ADC) can be used to quantify the electronic gain in the system in terms of ADC/mV. The right panel of figure \ref{fig:charge_inj} shows the distribution of this gain for the 32 channels of this ASIC. We can see that they vary by about $10\%$. It should be noted, however, that here the electronic gain (the High Gain Amplification from table \ref{tab:parameters}) was equal for all channels. This 8 bit value can be adjusted for each channel to modify the electronic gain by a factor of around 6. Therefore variations of the order of $10\%$ between channels can be removed through optimization if needed. 

\subsection{Electronic Noise}\label{sec:noise}

\begin{figure}[h!]
\centering
 \includegraphics[width=15 cm]{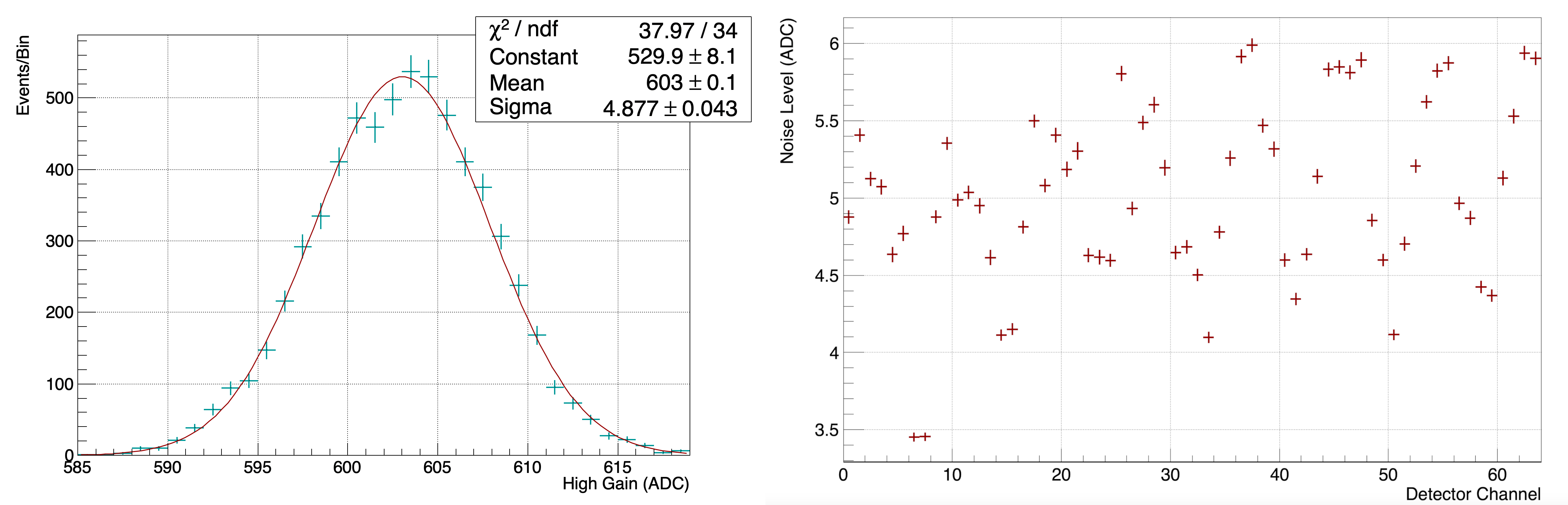}
 \caption{\textbf{Left:} The distribution of the high gain ADC values for one channel for 5000 forced readouts without any high voltage. The distribution is fitted with a Gaussian function where the width indicates the electronically induced noise. \textbf{Right:} The $\sigma$ values extracted from the Gaussian fits for the 64 channels on the readout system.}
 \label{fig:noise}
\end{figure}

To produce scientific data, the system will be triggered to perform a readout of all the ADC values when one, or potentially several, detector channels are above the charge or time threshold. However, in order to measure the baseline as well as the electronic noise, the system can be forced to trigger using the 'trig in' LVDS line. During standard data taking this can be done, for example, once per second to monitor potential drifts in the baseline due to temperature. Here we use this function to measure the electronic noise. For this purpose approximately 5000 triggers were forced while the high voltage was switched off, thereby only measuring the baseline, and its variations for the 64 channels, both in the high gain and low gain output. An example of the ADC distribution of channel 0, fitted using a Gaussian function can be seen on the left in figure \ref{fig:noise}. As no HV is applied to the SiPM, the noise seen here is that produced by the electronics on the front-end and injected into the HG line of this ASIC channel. The $\sigma$ values for the 64 channels, for the high gain, can be seen on the right. Similar values were found for the low gain, indicating a typical electronics-induced noise of the order of 5 ADC. 

The electronic noise can consist of two components, one which is shared between all the channels in an ASIC, or within the full board, and another which is individual for each channel. The presence of any shared noise component was tested by taking the mean ADC value of the 64 channels and studying its distribution. The resulting distribution was found to have a width consistent with 0 ADC, thereby excluding the presence of a significant shared noise component. The distribution in figure \ref{fig:noise} therefore shows the channel individual noise. It should be noted that the same level of noise was found by fitting the distributions in figure \ref{fig:charge_inj} acquired through direct charge injection.

The baseline can also be fitted from data when a high voltage is applied. Performing an accurate fit is more complicated in this situation, as the spectrum will also contain dark noise, not allowing for accurate automatic fits. However, several manual fits, using spectra such as that later discussed in figure \ref{fig:finger_fit}, provide similar values for the baseline width. This indicates that the HVPS does not induce additional noise when operational.

\subsection{Rate Capability}\label{sec:dead}

The SiPM readout is originally designed to observe transient sources like GRBs. As such it is designed to handle relatively high photon rates. From the POLAR mission, the brightest observed GRB was the short GRB 170127C which, after correcting for dead time, induced a gamma-ray rate of $\approx\,200\,\mathrm{ph\,cm^{-2}\,s^{-1}}$ for a duration of 100 ms. For the brightest GRB ever observed, 221009A the GRBAlpha collaboration found a flux of $1020\,\mathrm{ph\,cm^{-2}\,s^{-1}}$  in the 50-300 keV range \citep{GRB_alpha}. Using such a rate, and combining it with the predicted effective area of POLAR-2 from \cite{ESRF_paper}, we find that, to be capable of observing such transients without a significant amount of dead time, the instrument should be capable of handling trigger rates of hundreds of $\mathrm{triggers/(cm^2\,s)}$. Assuming the baseline design, which reads out 64 SiPMs with dimensions of $6\times6\,\mathrm{mm^2}$, this corresponds to being able to handle a trigger rate of the front-end board of the order of several kHz. 

The main sources which limit the readout rate of the system are identified as, the ADC readout time, the event packaging time in the FPGA and the readout rate of the board by the back-end. The 14-bit ADC readout is the first bottleneck and puts a lower limit on the possible time between two triggered events. With a clock speed of 1.5 MHz, the total readout time is $9.3\,\mu s$.

Each trigger creates a data packet, currently containing both the low and the high gain ADC values of the 64 channels, the trigger pattern and a time stamp. Processing this in the FPGA requires time, foreseen to be of the order of 20 $\mu s$. The trigger packets are stored in a buffer with a depth of 8, therefore inducing a second bottleneck around $150-200\,\mu s$. 

Finally, once the data packets have been prepared in the FGPA they need to be read out by a back-end through the 'data read' lines. This back-end system is independent of the readout discussed here. Two different back-ends were used for tests here. One which induces a readout delay of 2 ms at high rates. For the second, slightly faster one, this is at 1.2 ms, while other constants were identical for both systems.

\begin{figure}[h!]
\centering
 \includegraphics[width=15 cm]{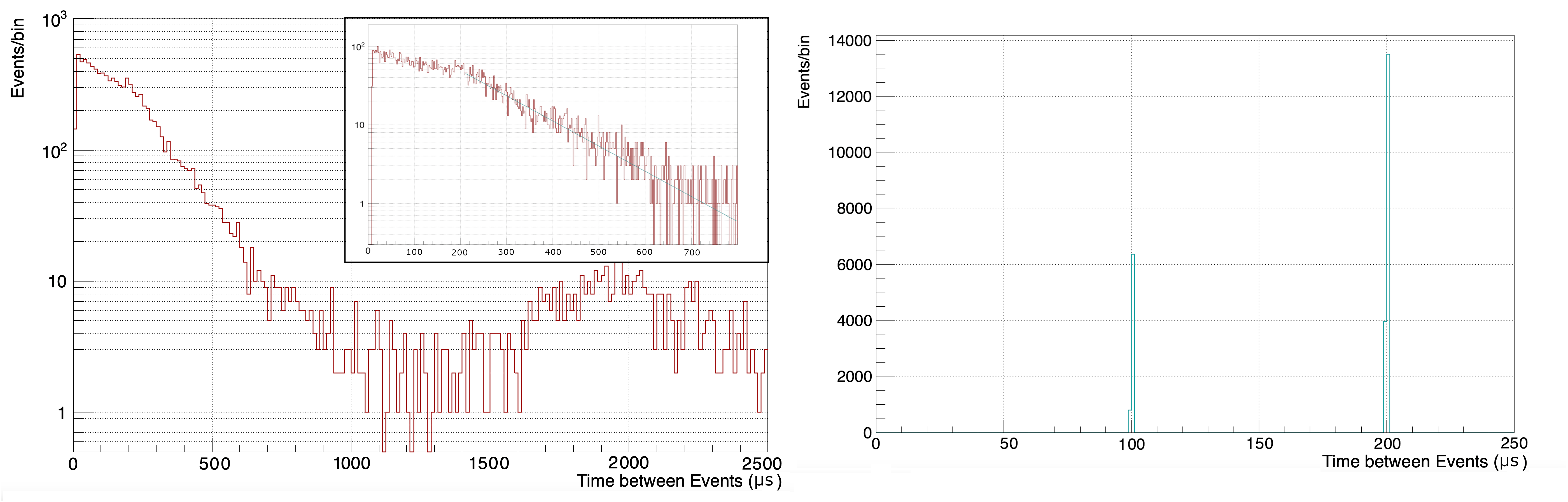}
 \caption{\textbf{Left:} The measured time between recorded events for a measurement of the dark noise. The inset shows the lower time region where the ADC readout-induced limit of 9$\mu s$ can be observed. \textbf{Right:} The measured time between recorded events for a measurement where a charge is injected at a constant rate of 10 kHz.}
 \label{fig:dead}
\end{figure}

The effects can be seen on the left in figure \ref{fig:dead} which shows the time between two consecutive events for a data run where thresholds were set such that the system was triggering at high rates on dark noise from the SiPMs. These measurements were taken at room temperature with new SiPMs. Typical dark count rates at the single p.e. level, based on the data sheets of these SiPMs, of several MHz are therefore expected. By setting the thresholds at 3 p.e. trigger rates of several 10's of kHz are produced. At large times, the peak induced by the back-end at 2 ms can be observed, while at lower times the effects due to the readout system itself. These can be more clearly observed in the inset. The lower limit at $9\,\mu s$ can clearly be observed, while a kink at around $200\,\mu s$, induced by the FGPA can be observed. Fitting the region below the kink using an exponential we can derive a trigger rate of 4 kHz. 

To further verify the trigger rate using a controlled trigger rate, charges were directly injected into the setup using the charge injection point on the PCB. The charges, produced by a pulse generator, are injected directly into the ASICs, thereby allowing to test the full electronics chain. When injecting charges at a rate of 1--2 kHz, the correct time between triggers could be recovered. Only when exceeding 5 kHz do we start to see this being distorted. The result of injecting charges at 10 kHz can be observed on the right of figure \ref{fig:dead}. While for an ideal detector, we would expect to measure 100~$\mu s$ delays, here, for about $60\%$ of the cases, a delay of 200 $\mu s$ is measured, indicating that packets are being lost. 

Overall, we can conclude that the system can handle rates of several kHz accurately, thereby allowing, when connected to a scintillator array with an effective area of $\approx10\,\mathrm{cm^2}$, to handle bright transient events, such as GRB 170127C, without significant dead time, while only for GRBs such as 220910A significant dead time will be incurred.

\subsection{Crosstalk}

A typical detector using scintillators, SiPMs, and electronics can suffer from various types of crosstalk. Firstly, there can be optical crosstalk between the scintillators. Secondly, crosstalk between microcells in the SiPM occurs. These first two forms of crosstalk depend on the detector design. The third, the electronic crosstalk, depends on the readout design. As the system discussed here is developed to read out 64 independent SiPMs, it is important to avoid any electronic crosstalk between readout lines in the system. A first method to check for electrical cross talk is through the use of injecting charges directly into the PCB. From the charge injection point, the charges go into the ASICs, thereby allowing to check for electrical crosstalk in the ASIC, ADC and in traces towards the FPGA. Even when injecting charges which saturate the LG no signals in adjacent channels were found. 

\begin{figure}[h!]
\centering
 \includegraphics[width=15 cm]{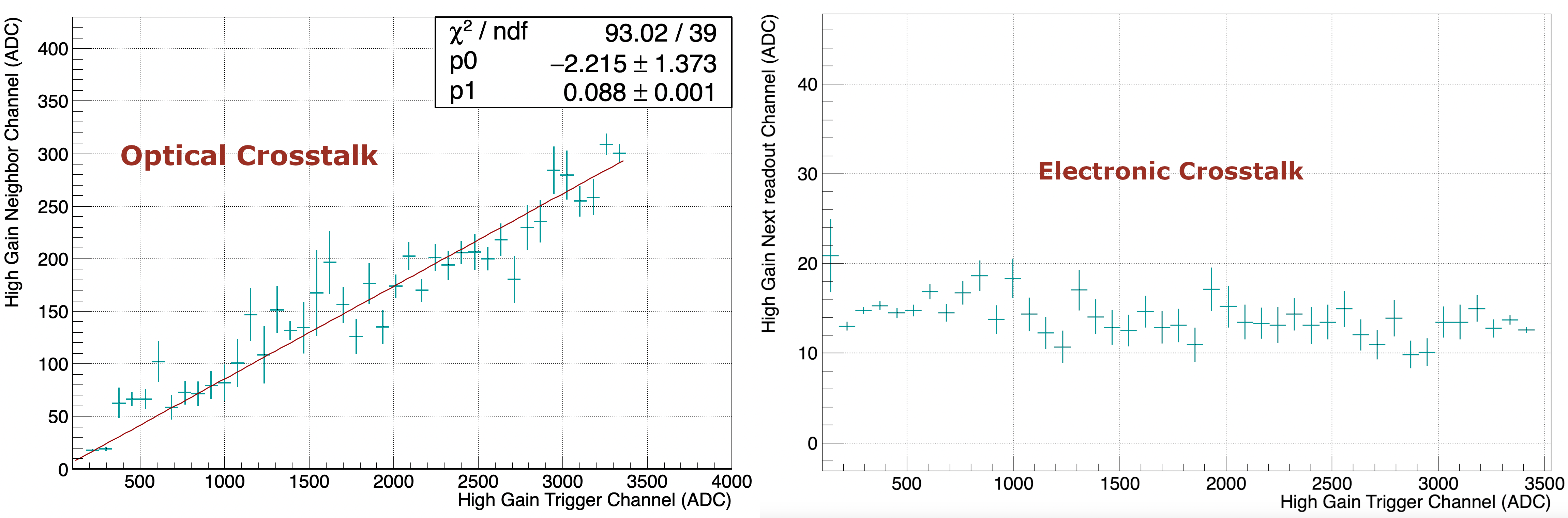}
 \caption{\textbf{Left:} An example of an optical crosstalk measurement acquired using an array of GAGG scintillator crystals that are poorly insulated. The correlation is fitted with a linear function which indicates a crosstalk of the order of $10\%$. \textbf{Right:} The correlation of the ADC value for two channels that have their readout lines adjacent, thereby potentially allowing for electrical crosstalk. No correlation is found, indicating that the system does not suffer from any significant electrical crosstalk.}
 \label{fig:xtalk}
\end{figure}

Although positive, this method does not exclude any electrical cross talk from the SiPM towards the ASIC. To verify that no significant crosstalk of this type exists also here, data with large signals from the SiPM can be used. For this purpose, we use data from acquisitions where the system was coupled to an array of GAGG crystals which was illuminated by a 100 keV beam. The correlation between ADC values in different channels was evaluated and studied for crosstalk. To verify that the analysis works, we initially looked at data from two SiPM channels which are geometrically adjacent. The mean ADC value as registered in a geometrically neighboring channel, as a function of the ADC value in the triggering channel, which here was illuminated by the beam, can be seen on the left in figure \ref{fig:xtalk}. We see a clear correlation between the two ADC values, indicating that a significant level of crosstalk, either optical or between SiPM channels, exists. This was expected, since the GAGG crystal array is not well optically insulated, thereby resulting in optical crosstalk of $\approx10\%$. 

Such a correlation was only observed for channels which are geometrically adjacent. However, we can also study two channels that are geometrically far apart but have adjacent lines from the ASIC to the ADC. Such a channel is highly unlikely to have optical crosstalk, while having a large probability for electronic crosstalk. An example of the correlation in ADC values for this case is shown on the right in figure \ref{fig:xtalk}. We can see no sign of correlation, thereby indicating that there is no significant electrical crosstalk. It should be noted that the mean ADC value for the electrically neighboring channel is not at 0: this is the result of dark noise in the SiPM.  

\section{Performance with EJ-248M and MPPC S13361}\label{sec:plastic_perf}

The SiPM readout system was originally designed as part of the POLAR-2 mission, a large-scale plastic scintillator-based polarimeter \citep{ESRF_paper}. It was therefore initially optimized to read out a densely packed array of plastic scintillator bars. Although originally, EJ-200 was chosen for this purpose, it was found that EJ-248M performed better in terms of light yield despite its lower intrinsic scintillation light yield compared to EJ-200. This was found to be a result of EJ-248M being a relatively harder plastic and therefore allowing a smoother surface quality, resulting in a larger fraction of the light reaching the SiPM. The details of this are discussed in \cite{POLAR-2_optical}, where the details on the optical coupling are also presented along with the calculation of the achieved light yield. In an optimized setup an average of $\approx 1.6$~p.e./keV was measured. 

Similarly, the combination of the readout system together with an EJ-248M, as well as an EJ-200 array, was tested during a beam test at the European Synchrotron Radiation Facility (ESRF) of which the details are presented in \cite{ESRF_paper}. While the details on the light yield and spectra can be found in \cite{ESRF_paper}, it should be noted that those results were produced using the same electronics, though with a slightly faulty firmware. Here, we present the typical response of the system after a firmware update. For details on the performance of the system with a polarized beam, we refer the reader to \cite{ESRF_paper}, while for details on the optical performance for this system, we refer the reader to \cite{POLAR-2_optical}.

Such measurements, as well as those discussed in this and the next section, were performed with an over-voltage of 3.0~V for all 64 channels. To achieve this, the breakdown voltages of all 64 SiPM channels were measured at various temperatures prior to mounting them on the PCB. This over-voltage was selected as it allows for photon detection efficiency of $\approx50\%$ while keeping the SiPM cross talk below $10\%$ \footnote{\url{https://www.hamamatsu.com/content/dam/hamamatsu-photonics/sites/documents/99_SALES_LIBRARY/ssd/s13360_series_kapd1052e.pdf}}. The POLAR-2 mission relies for its scientific measurements on the detection of coincident energy depositions in 2 detector channels, one of these energy depositions can be as low as several keV. For this mission it is therefore important to keep the energy threshold as low as possible. As this threshold is driven by a combination of the dark noise and the crosstalk, a low crosstalk in the SiPM is more important than a large photon detection efficiency. For a spectrometer with less readout channels, one might consider increasing the over-voltage to prioritize the photon detection efficiency over a higher crosstalk. 

\subsection{Spectral Measurements}

\begin{figure}[h!]
\centering
 \includegraphics[width=12 cm]{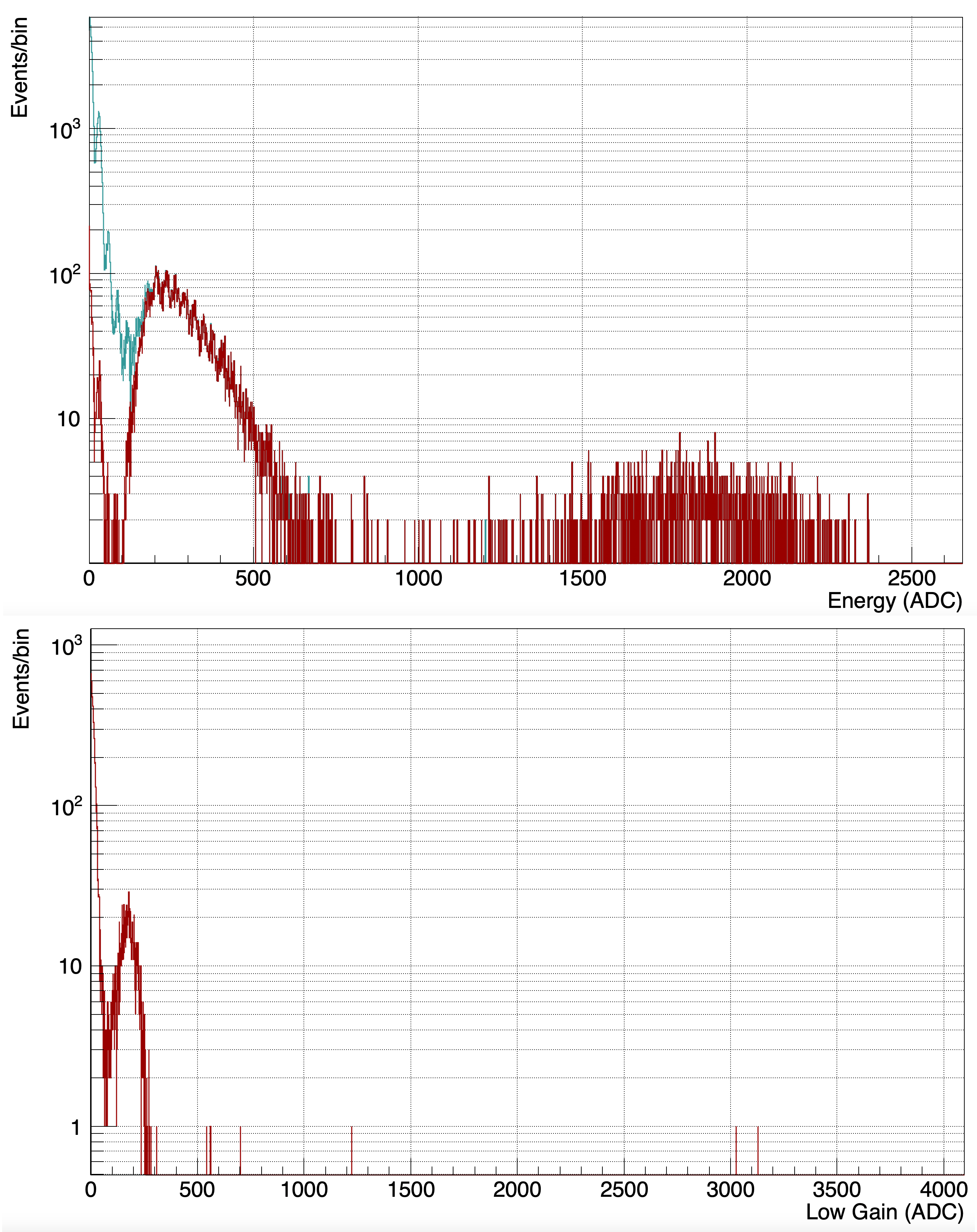}
 \caption{The energy spectra, in ADC, after pedestal subtraction both in the high gain range (top) and low gain range (bottom). These spectra were obtained when irradiating  the readout system connected to a plastic scintillator array at room temperature with an $^{241}Am$ source. The events that induced a charge trigger are shown in red, while those that are under the threshold are shown in green. The photo-absorption peak can be seen at around 1800 ADC, while the Compton-dominated spectrum can be seen at lower energies. Below $\approx400\,\mathrm{ADC}$, the individual photo-electron peaks can be clearly observed.}
 \label{fig:spectrum_plastic}
\end{figure}

In order to study the performance of the readout system with an array of EJ-248M scintillators, a full detector, consisting of 64 scintillator bars, was scanned using a collimated $^{241}Am$ source. The emission spectrum of this source is dominated by a $59.5\,\mathrm{keV}$ line. The spectrum from one of the detector channels can be seen in figure \ref{fig:spectrum_plastic}. In this spectrum, which is shown after pedestal subtraction was performed, events that have induced a charge trigger are shown in red, while those that did not are shown in green. It can be seen that the threshold was set so low that a small fraction of events from the dark noise region induced a trigger. This is possible as the threshold has a finite width resulting in a non-zero threshold efficiency at low ADC values. This, in combination with the very high rates in the dark noise region, results in such triggers. Increasing the threshold by approximately 50~ADC fully removes the dark noise-induced triggers.

Within the detector used here, the $59.5\,\mathrm{keV}$ photons are absorbed in a plastic scintillator which, based on its specifications, converts this into approximately 550 optical photons. Based on simulations presented in \cite{POLAR-2_optical} about 1/3rd of these optical photons reach the SiPM where, as a result of the photon detection efficiency of the SiPM $50\%$ of these are converted to photo-electrons which, the electrical signal from these gets amplified and finally converted into a digital signal shown on the x-axis of figure \ref{fig:spectrum_plastic}. Overall, based on the various conversions one expects about 1 p.e. per keV deposited. The charge trigger applied here corresponds to approximately 5 p.e., or 5~keV. The individual photo-electron peaks can be clearly observed up to more than 10~p.e.. In addition, the photo-absorption peak resulting from 59.5~keV absorption in the plastic can be seen around 1800~ADC. The spectrum, as observed in the low gain range also shows this peak, around 180\,ADC, indicating that the dynamic range in the low gain reaches up to approximately 1.3 MeV. This is achieved by setting the HG and LG Gain DAC values equal. The range can be optimized for specific purposes by changing these values. 

\subsection{Non-linear Gain \label{sec:non_linear}}

\begin{figure}[h!]
\centering
 \includegraphics[width=15 cm]{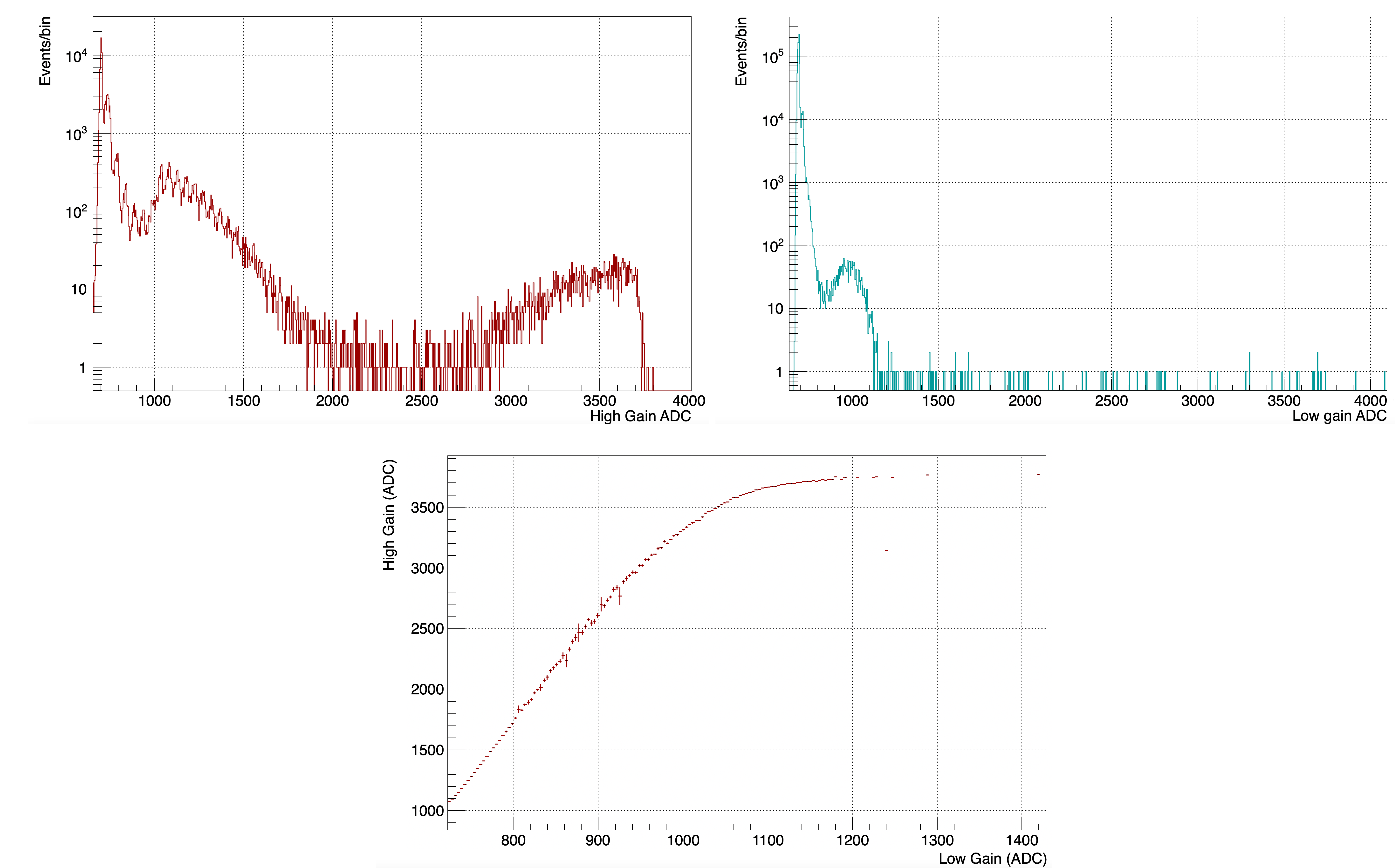}
 \caption{The energy spectra, in ADC, here before pedestal subtraction both in the high gain range (top left) and low gain range (top right). These spectra were obtained when irradiating  the readout system connected to a plastic scintillator array at room temperature with an $^{241}$Am source. The photo-absorption peak can be seen at around 3500~ADC in the high gain and 900~ADC in the low gain. However, we see that the peak is cutoff in the high gain due to saturation. The bottom panel illustrates the saturation induced non-linearity further by plotting the high gain ADC value versus those of the low gain.}
 \label{fig:non_linearity}
\end{figure}

When calculating the dynamic range in the previous section, we assumed a fully linear relation between keV and ADC. This is, however, not correct due to saturation of the amplifiers in the Citiroc 1A ASIC. This has previously been discussed in \cite{ESRF_paper}. This saturation can be noticed in a spectrum which is not fully contained within the dynamic range, or when using the direct charge injection to produce a figure such as figure \ref{fig:charge_inj}. Figure \ref{fig:non_linearity} shows the (non-pedestal subtracted) high gain (top left) and low gain (top right) for a second channel from the same measurement as used for figure \ref{fig:spectrum_plastic}. Whereas in figure \ref{fig:spectrum_plastic} the photo-peak is fully contained, here the photo-peak can be clearly seen to be cut off in the high gain spectrum which shows no events beyond 3800~ADC whereas the ADC range extends up to 4095~ADC. In the low gain the peak is fully visible, while, in addition, cosmic ray induced events can be seen up to 4095~ADC. When plotting the high gain against the low gain ADC values for this measurement, the curve shown in the bottom panel of figure \ref{fig:non_linearity} is achieved. The high gain ADC value is linear with the low gain, and therefore with the keV input, up to approximately 2000~ADC. Above this, the amplifier in the ASIC starts to saturate, until at 3100~ADC it is fully saturated. As a result, no ADC values exceeding this value are observed. This indicates that the saturation of the amplifier causes $25\%$ of the dynamic range in the high gain to be lost. However, above this range one can rely on the low gain ADC values where no such saturation is observed. 

It should further be noted that for the channel studied here, the dynamic range significantly differs from that of the channel studied in the previous section.  This is a result of a difference in the gain of the two channels. Both channels are readout with SiPM channels which are operated at equal over-voltages\footnote{The breakdown voltages of the various SiPM channels and their temperature dependence were calibrated prior to mounting them on the PCB}.  We do, however, observe that for the channel used here we have a gain of approximately 49 ADC/p.e., whereas for the channel of figure \ref{fig:spectrum_plastic} the gain is 25~ADC/p.e.. This gain can be adjusted using the 6 bit DAC of the high gain which was here set to be equal for all channels. In case a higher uniformity between the various channels is desired, this DAC value can be used to equalize the gains. This has previously been achieved for POLAR-2 and a detailed discussion of this is presented in \cite{NDA_thesis}. For further information, which requires details on the calibration of POLAR-2: the full calibration of one POLAR-2 module, and the resulting distribution of the gain from various channels is discussed in detail in \cite{NDA_thesis}.

\subsection{Photo-electron Resolution}

\begin{figure}[h!]
\centering
 \includegraphics[width=12 cm]{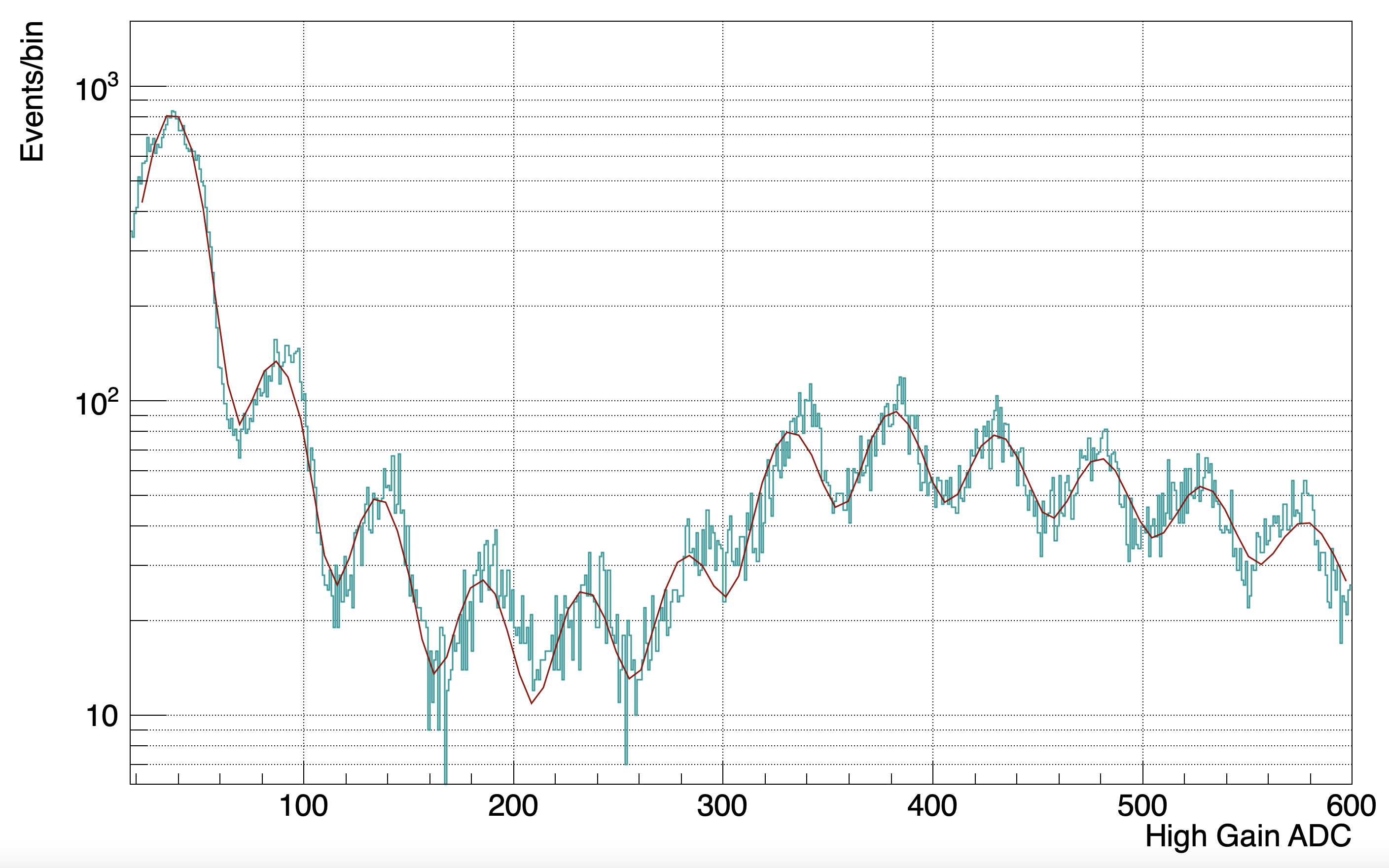}
 \caption{The low energy part of the high gain ADC spectrum fitted using a sum of 12 Gaussian functions where the distance is fixed and the width is described according to $\sigma_{p.e} = \sqrt{\sigma_1^2 N_{p.e.}+ \sigma_e^2}$.}
 \label{fig:finger_fit}
\end{figure}

Both the spectra of figures \ref{fig:spectrum_plastic} and \ref{fig:non_linearity} show numerous photo-electron peaks, also called fingers in the field of SiPMs. The capability to observe and fit numerous fingers allows to measure the number of photo-electrons produced per deposited keV and therefore study the photo-detection efficiency of each channel. Subsequently, one can fine tune the bias voltages for each channel, as well as the electronic gain to produce a uniform sensitivity in all the detector channels.  Within one channel, each finger in the spectrum corresponds to one photo-electron and therefore on fired microcell in the SiPM. The distance between the fingers should be constant. In addition, the width of the fingers increases with their number \citep{NDA_masterthesis} according to:

\[\sigma_{p.e} = \sqrt{\sigma_1^2 N_{p.e.}+ \sigma_e^2}\]

\noindent where $\sigma_{p.e}$ is the width of the finger retrieved from a Gaussian fit, $N_{p.e.}$ is the number of the finger and $\sigma_e$ is the width of the pedestal electronic noise. The latter contains the width of the electronics induced noise which can be measured using the width of the pedestal as presented in section \ref{sec:noise}. However, whereas the pedestal only contains the noise induced by the electronics, the zeroth finger also contains electronic noise from the SiPM itself, causing $\sigma_e$ to be larger than the noise measured through the pedestal. $\sigma_1$ represents the increase in the width due to amplification of the noise. Using this relation for the width of the fingers, leaving the height of the various fingers as a free parameter, and using a single fitting parameter for the width between peaks, we can fit the first 12 fingers visible in the high gain spectrum of Figure \ref{fig:non_linearity}. The result can be seen in figure \ref{fig:finger_fit}, i.e. the fingers follow the expected behaviour with a fixed distance of 49~ADC, a $\sigma_e=11.9$ ADC and $\sigma_1=3.4$ ADC. Using the position of the photo-peak, we know 49~ADC corresponds to approximately 1 keV. Therefore we know that $\sigma_e=0.24$ keV and $\sigma_1=0.07$ keV. The ratio of $\sigma$ over the finger separation observed here is primarily a result of the choice of SiPM. Using SiPMs with a smaller terminal capacitance such as those with smaller channels, rather than the $6\times6\,\mathrm{mm^2}$ ones used here, will result in smaller widths. Initial tests of the same front-end with Hamamatsu SiPMs of the S14 type deemed to be unsuccessful as this type has a larger terminal capacitance, thereby making it almost impossible to see the fingers in the spectrum \cite{NDA_thesis}.

\section{Performance with GAGG and MPPC S13361}\label{sec:GAGG_perf}

\subsection{Test Setup}

\begin{figure}[h!]
\centering
 \includegraphics[width=8 cm]{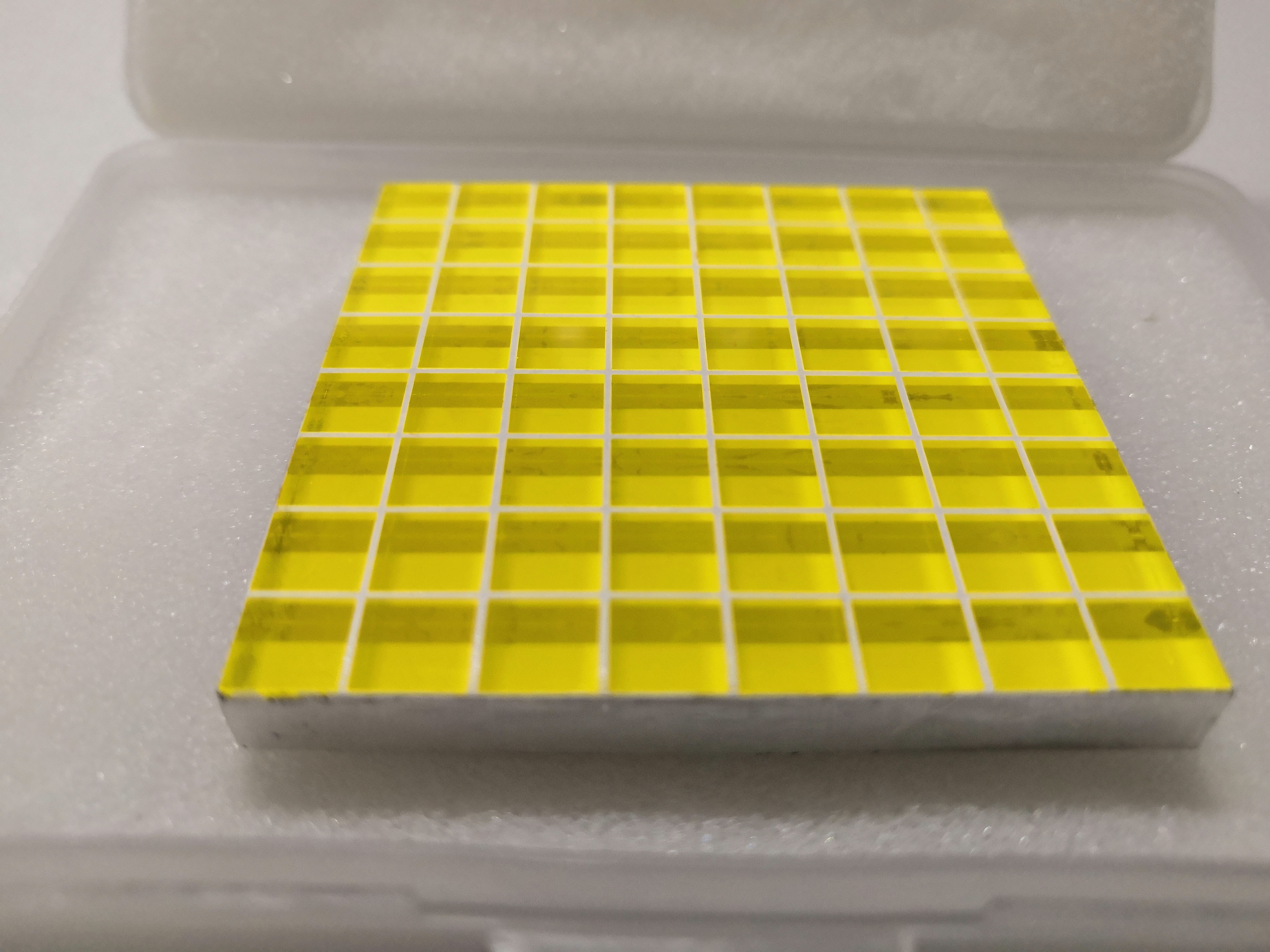}
 \caption{The GAGG array used to study the behavior of the system at the LARIX-A X-ray facility. The GAGG crystals match the $6\times6\,\mathrm{mm^2}$ area of the SiPMs and have a height of $15\,\mathrm{mm}$. They are separated by a thin layer (white) of barium sulfate.}
 \label{fig:GAGG_array}
\end{figure}

The SiPM readout system was additionally tested for use with a high-Z scintillator. To date, the system has been tested using both CeBr$_3$ and GAGG:Ce. For CeBr$_3$ only basic tests were performed which showed that the scintillator can be read out successfully. The combination of the system together with an array of 64 GAGG crystals was tested in more detail using the LARIX-A X-ray facility of the University of Ferrara \citep{Virgilli2014}. Similar to the tests with plastic scintillators described in section \ref{sec:plastic_perf}, the system was equipped with 4 arrays of $4\times4$ MPPCs from Hamamatsu of the S13360 type, the S13361-6075PE-04. The array of GAGG scintillators consists of crystals with dimensions of $6\times6\times15\,\mathrm{mm}^{3}$ which are placed in a barium sulfate (BaSO$_4$) housing which provides optical decoupling of the crystals in the array as well as mechanical support. Figure \ref{fig:GAGG_array} shows the array, which is coupled directly to the SiPM array shown in figure \ref{fig:thermomechal_design}. A 1 mm thick carbon fiber cover was placed over the array to provide light tightness and to couple it mechanically to the frame. The combination was placed such that the collimated beam is perpendicular to the array.

The crystals are arranged such that they couple exactly to the SiPM array placed on the readout system. Finally, the GAGG array is mechanically coupled to the readout system using a custom-made carbon fiber housing, which ensures a light-tight readout system as well as mechanical stability.

The detector produced in this way was placed in the LARIX X-ray beam such that the beam (with a size of $7\times7\,\mathrm{mm^2}$) entered the detector from zenith, meaning the beam was perpendicular to the detector array. As the size of the beam matches the size of the crystals, the system could be aligned, using an x-y table, such that the beam entered primarily a single crystal, while some neighboring crystals were partially irradiated as well. Finally, the beam energy was changed within a range between 25~keV and 75~keV. 

\begin{figure}[h!]
\centering
 \includegraphics[width=15 cm]{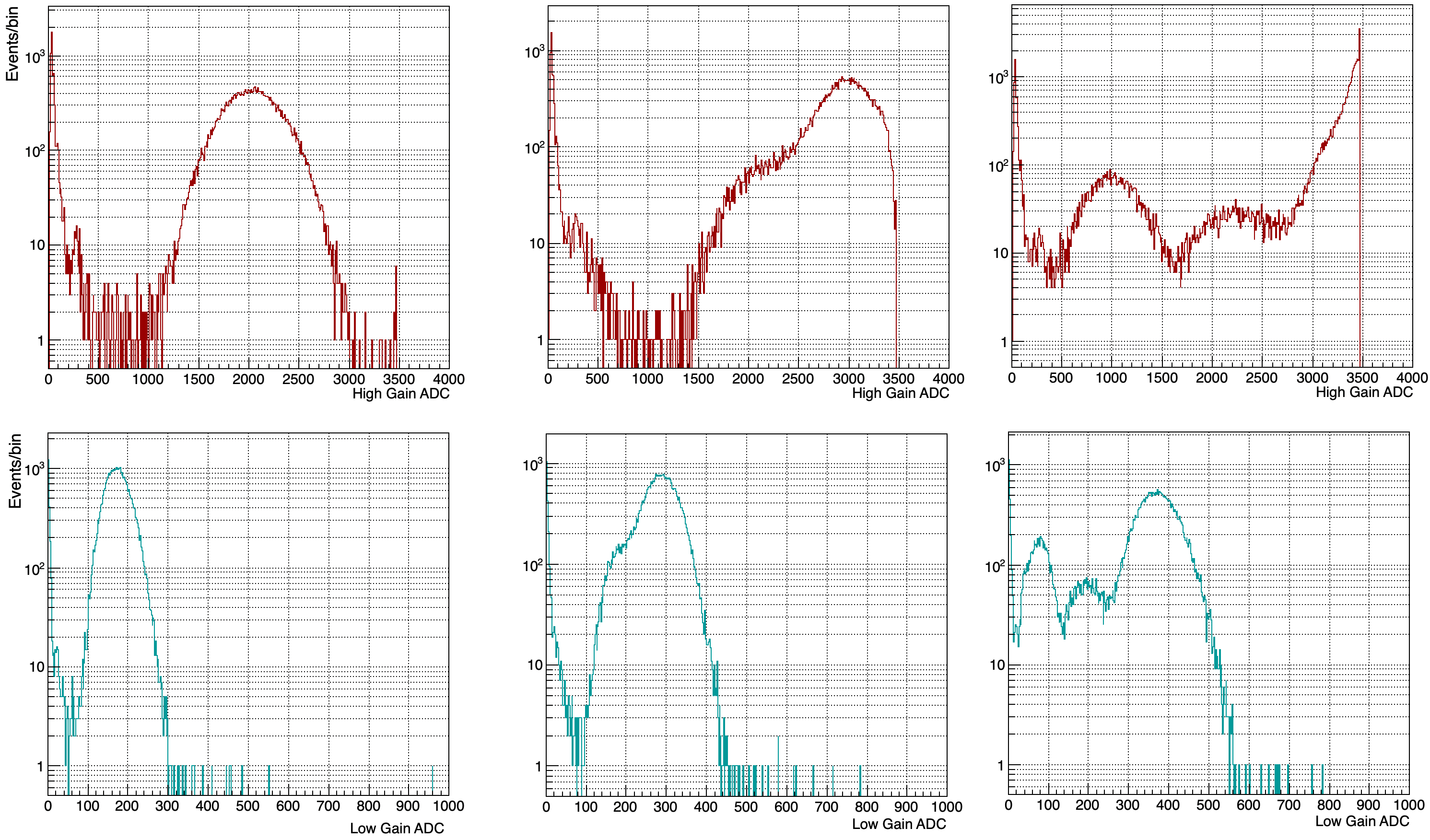}
 \caption{The spectra as measured using a single channel for both the high gain (top) and low gain (bottom) for a 30~keV (left), 45~keV (middle) and 60~keV (right) beam. The 30~keV high gain spectrum shows a clear photo-peak resulting from the 30~keV photons. The 45~keV spectrum is more complex due to the presence of a fluorescence line at 32~keV from Barium (from the BaSO$_4$ housing). The spectra resulting from the 60~keV beam are not fully visible in the high gain range, while in the low gain 3 peaks are visible: the 60~keV peak, the 32~keV fluorescence from Barium due to the BaSO$_4$ housing, and the 17~keV escape peak. The latter two are also visible in the high gain spectrum. }
 \label{fig:spectra_GAGG}
\end{figure}

\subsection{Spectral Measurements}

Examples of pedestal subtracted spectra measured using a single channel for 30, 45 and 60~keV are shown in figure \ref{fig:spectra_GAGG}. The spectra in figure \ref{fig:spectra_GAGG} show a range of different photo-peaks. The spectrum from the 30~keV beam only shows the 30~keV peak, which is around 2000~ADC in the high gain output and at around 173~ADC in the low gain output. For the 45~keV beam, the 45~keV photo-peak is visible, although somewhat distorted through non-linearity in the high gain. In addition, this peak has a shoulder at around 2200 ADC resulting from a fluorescence line that corresponds to 32~keV and was identified as the Barium $K_\alpha$ fluorescence line from the BaSO$_4$ housing. This line is also clearly visible in the 60~keV spectrum, both in the high (2200 ADC) and low gain (190 ADC) data. In the low gain spectrum, the 60~keV photo-absorption peak is also clearly visible, while finally, a 3rd peak is visible at 1000 ADC in the high gain and at around 80 ADC in the low gain. This peak is the escape peak resulting from 60~keV photons which produce one 43~keV fluorescence photon through absorption in the Gallium of the GAGG which escapes the crystal, while the remaining energy is measured. This peak was also visible at higher beam energies and moved as expected (e.g. up to 1850 ADC in high gain for a 70~keV beam, corresponding to 26~keV). The 32~keV line from Barium becomes less dominant with increasing energy as the BaSO$_4$ becomes more and more transparent at higher energies.

\subsection{Energy Resolution}

\begin{figure}[h!]
\centering
 \includegraphics[width=13 cm]{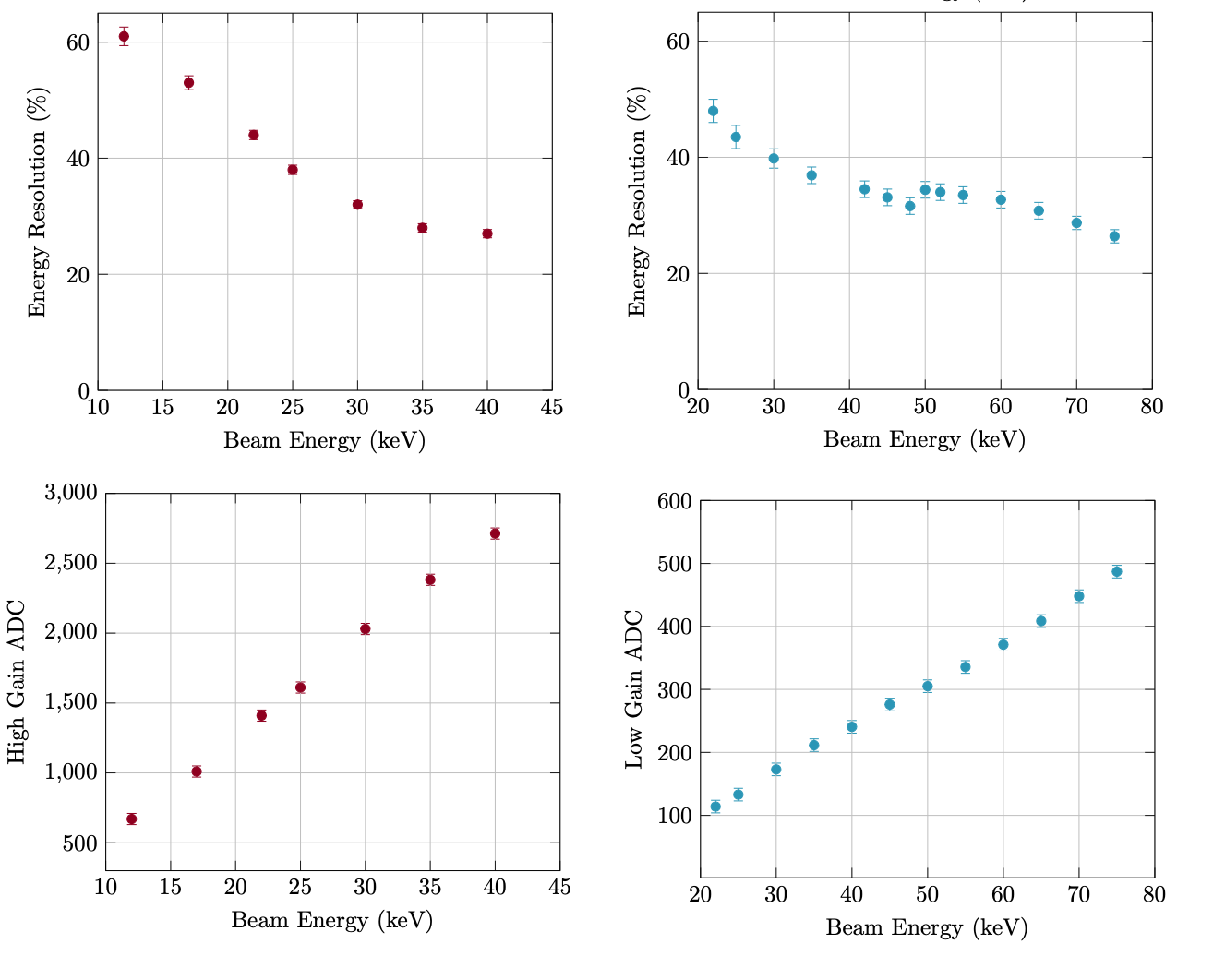}
 \caption{The energy resolution (in FWHM) as measured using both the pedestal subtracted high (top left) and low (top right) gain output as a function of energy. The bottom two figures show the correlation between the measured mean of the photo-peak in ADC, left for high gain and right for low gain, vs. the beam energy in keV. 
 }
 \label{fig:energy_res}
\end{figure}

Using the various features in these measured spectra we can look at the energy resolution, defined here as the full-width half maximum (FWHM) over the mean, as a function of energy. This can be measured both using the high gain, for energies up to 40~keV above which the gain becomes non-linear, as well as using the low gain starting from 20~keV at which point the absorption peaks are significantly separated from the noise region. The results, based on fits of both photo-peaks and escape peaks, can be seen in figure \ref{fig:energy_res} which shows the resolution measured in low gain on the right and high gain on the left. It can be seen that at low energies the energy resolution in the high gain is significantly better than that measured in the low gain, indicating deterioration of the energy resolution at low ADC values in the low gain. Around 40~keV the two resolution values become comparable. The K-edge from Gallium, at 50~keV, starts to become clearly visible. Above the K-edge it becomes significantly more likely that the energy is deposited by two individual interactions, one from the K-edge absorption and one from the remaining photon energy, rather than one. This deteriorates the energy resolution. The energy resolution achieved here, which was measured without optimizing the setup for this purpose, is comparable to those mentioned in for example \cite{energy_res}, where an energy resolution of $\approx45\%$ is presented for 22~keV and $\approx25\%$ is presented for 88~keV. As will be discussed next in section \ref{sec:Dyn_range}, the optical coupling between the GAGG and the SiPM can likely be improved significantly, as here no optical coupling was used. This would result in more optical photons/keV reaching the SiPM and therefore in a better energy resolution. 

Although the energy resolution looks good at the lower energies tested here, it would be useful to also irradiate this setup with higher energy photons. For this detailed tests with both radioactive sources as well as synchrotron beams with energies up to $500\,\mathrm{keV}$ will be performed in the future.

\subsection{Dynamic Range}\label{sec:Dyn_range}

Figure \ref{fig:energy_res} in addition shows the correlation between the ADC value and the input energies extracted from the measured spectra. A good linearity can be observed. We can extrapolate the dynamic range of the low gain output by assuming a linear behavior up to 4095 ADC, up to $\approx640\,\mathrm{keV}$. It is important to note, however, that for these tests no optimization was performed to maximize the dynamic range, which can easily be extended by changing the electronic low gain (which here was set equal to the electronic high gain DAC). By varying the values of both, an optimization can be performed for a specific overlapping range for both outputs or to extend the low gain output beyond 640~keV. 

\begin{figure}[h!]
\centering
 \includegraphics[width=10 cm]{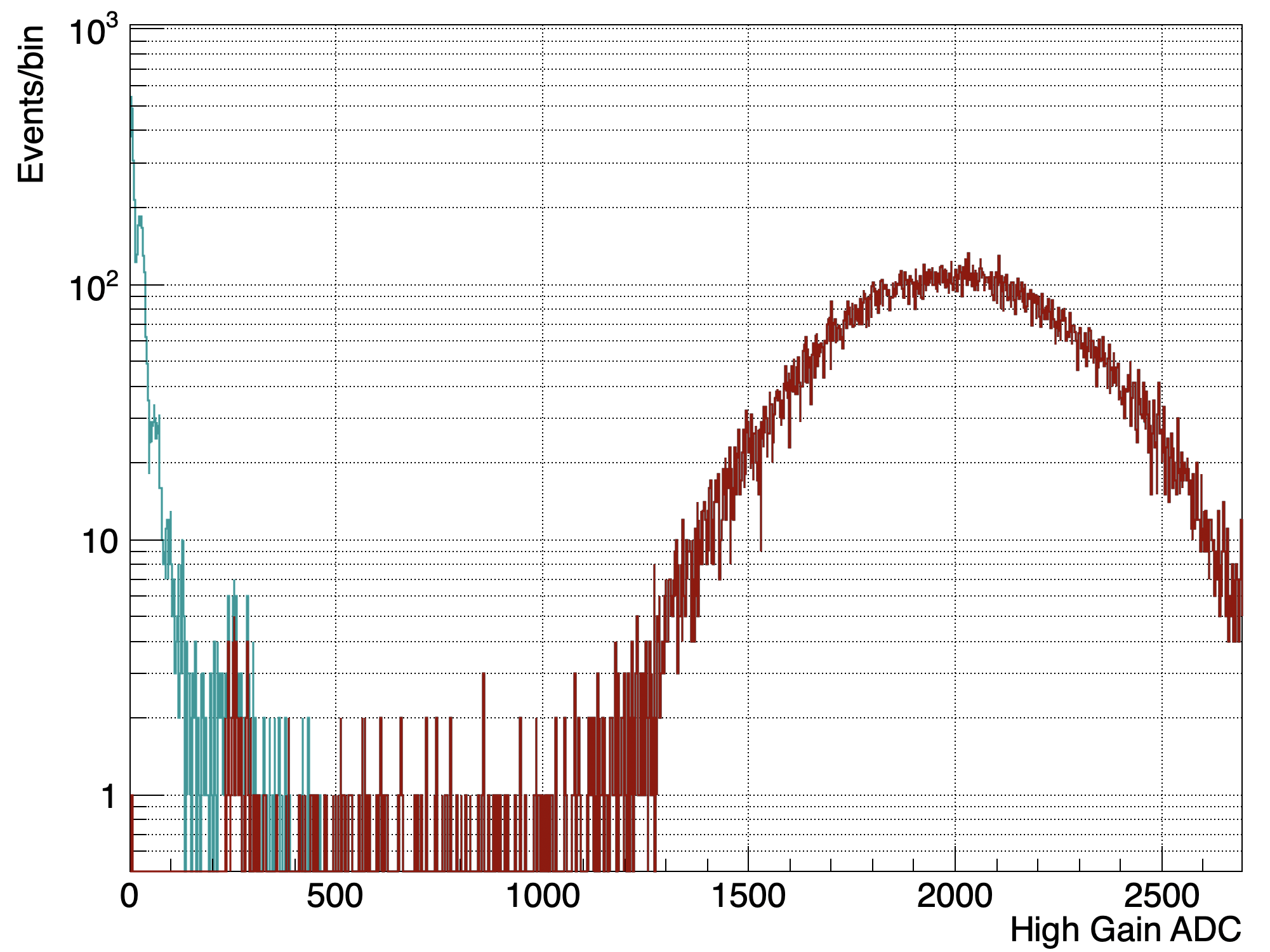}
 \caption{A detailed view of the high gain spectrum achieved using a 30~keV beam with the GAGG setup. The hits which induced a charge trigger are shown in red, while the others, which result from a trigger in one of the other channels are shown in green. The trigger threshold corresponds to 250~ADC, which corresponds to 3.75~keV or 8 photo-electrons. The first 4 photo-electrons can be observed below the trigger threshold, beyond the 4th the statistics is too low to see these.}
 \label{fig:trigger_sepc}
\end{figure}

While the results from figure \ref{fig:energy_res} indicate that a dynamic range up to 640 keV can be achieved relatively easily, the lower bound of the dynamic range can be studied by looking at the threshold position in the high-gain ADC spectrum. For the tests performed here, only the charge threshold was used by setting the time threshold to its maximum (the DAC was set to 1024 for both ASICs). The overall charge threshold was then set such that without the X-ray beam irradiation, the trigger rate was of the order of several Hz. No fine-tuning of the individual channels using the 4-bit DACs was performed. The threshold position can be seen in figure \ref{fig:trigger_sepc} where all events which induce a charge trigger are shown in red, while those induced by triggers in other channels are shown in green. The threshold position, which was measured by dividing the triggered spectrum by the spectrum containing all events and fitting the result with an error function, was found to be 250 ADC. Using the position of the 30~keV peak, this can be translated to be 3.75~keV. In addition, by fitting the distance between the fingers, which can be observed at low ADC values, we can deduce a light yield of 2.3~p.e./keV. Using this we can finally calculate the threshold to be equivalent to 8.3 photo-electrons. 

It should be noted that the light yield of 2.3~p.e./keV is rather low for such a setup. This is a result of the degraded quality of SiPMs used in this test, as well as a lack of good optical coupling. The SiPM arrays used here were already once unsoldered from a previous setup, during this process the heat has caused the protective resin to darken, thereby decreasing the light collection efficiency. In addition, no grease or optical coupling was used in between the GAGG and the SiPM. Based on measurements with the EJ-248M plastic scintillators, where a light yield of 1.6~p.e./keV could be achieved, and the scintillation yield of GAGG being about 5 to 6 times higher than EJ-248M, we can assume that through optimization a significantly higher light yield, of at least 3~p.e./keV should be achievable.

While such a trigger threshold is already very good for such a spectrometer (given the simple optical setup), it should be noted that the measurements were also performed without any cooling while having the electronics fully contained inside the mechanics, thereby not allowing air to circulate. As a result, the SiPM arrays were at a temperature of $28^\circ$ C during data taking. At room temperature one could therefore lower the threshold by approximately 1~p.e. while keeping the same rate, thereby reaching close to a threshold of 3~keV. Furthermore, as mentioned before, the optical coupling between the GAGG and the SiPM array was not optimal at this test, therefore it is likely that the light yield can be further improved.

\subsection{Further potential optimization}

The purpose of the tests at the LARIX facility was to verify that the system can be used successfully to read out scintillators which are slower than plastic. The results indicate that using this simple setup, GAGG can be read out with an expected energy resolution within an energy range of 3.75 to 640~keV. This range can, however, be easily adapted to optimize the system for a specific purpose. As mentioned earlier, the energy threshold can be lowered by improving the optical coupling, or by simply lowering the readout temperature. The range can furthermore be changed by varying the high and low gain DAC values. For example, by reducing the overlap between the high and low gain range, the high energy limit can be increased beyond 640~keV.

The 640~keV calculated here assumes a fully linear relation between the input energy and the measured ADC value. We know already, as discussed in section \ref{sec:non_linear}, that for the high gain the amplifier is not fully linear at high ADC values resulting in some saturation. In addition, saturation of the number of microcells in a SiPM array will also cause non-linearity, thereby limiting the dynamic range. We can calculate the effect of saturation of the used SiPM type, the S13361-6075, as well as that for the versions with $50\mu m$ and $25\mu m$ cells. For this we use the following relation \citep{DeAngelis_2023}:

\[N_{p.e.}=N_{microcell}(1-e^{-N_{opt.ph.}PDE(1+\mu)/N_{microcell}})\]

\begin{figure}[h!]
\centering
 \includegraphics[width=12 cm]{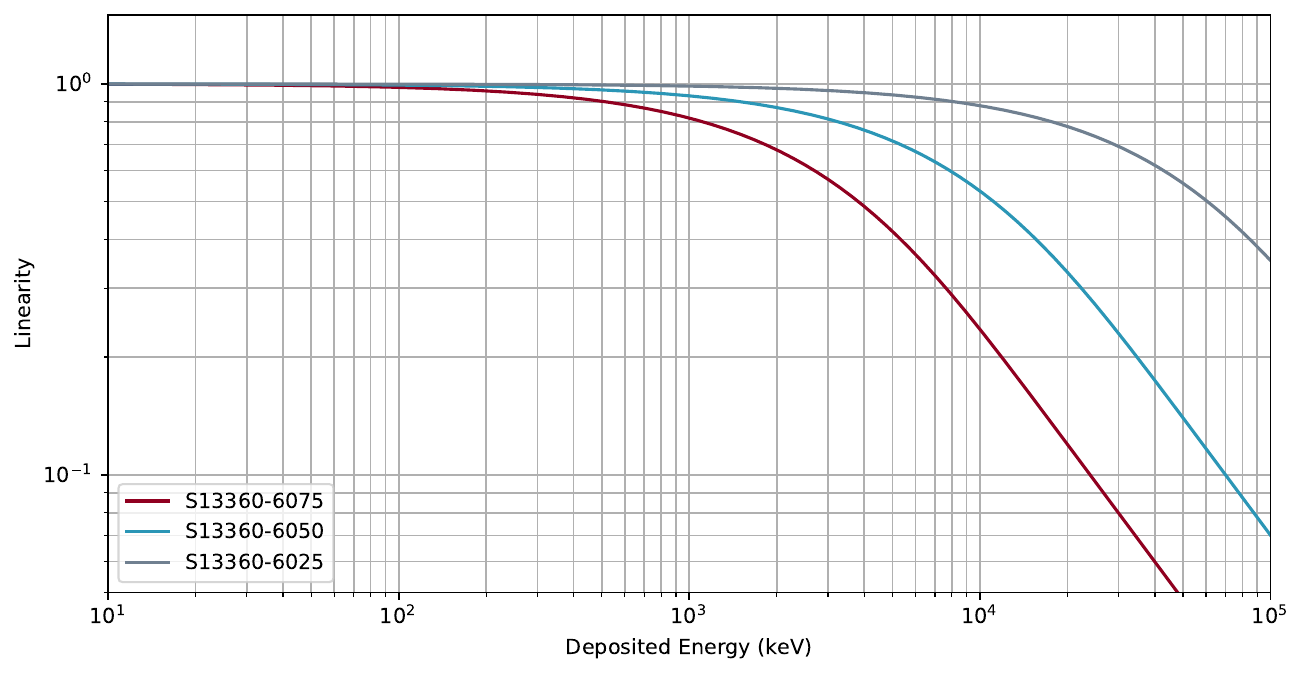}
 \caption{The linearity, computed as the relative signal height produced by a SiPM over that from a hypothetical SiPM with an infinite number of microcells for 3 different types of SiPMs coupled to a GAGG crystal as a function of energy. It can be seen that while the S13361-6075 starts to show significant deviations from a linear behavior above 500~keV, the S13361-6025 shows linear behavior well into the MeV region.}
 \label{fig:linearity}
\end{figure}

Here $N_{p.e.}$ is the number of detected photons (photo-electrons, or number of fired microcells), $N_{microcell}$ the total number of microcells in the SiPM, $N_{opt.ph.}$ the number of optical photons arriving at the SiPM, $PDE$ the photo-detection efficiency and $\mu$ the crosstalk. The values for $\mu$, $PDE$ and $N_{microcell}$ are all taken from the Hamamatsu data sheet\footnote{\url{https://www.hamamatsu.com/content/dam/hamamatsu-photonics/sites/documents/99_SALES_LIBRARY/ssd/s13360_series_kapd1052e.pdf}}, while the $N_{opt.ph.}$ was converted to keV using the measured light yield of 2.3 p.e./keV and the typical scintillation yield of GAGG (60 optical photons/keV). The result of this is shown in figure \ref{fig:linearity}, which shows the relative gain for the three different types of SiPM as a function of energy. We can see that for the S13361-6075 used here, the SiPMs start to show a non-linearity exceeding $10\%$ for energies above 500~keV while measuring anything above 1 MeV becomes challenging. For the GAGG system, the use of S13361-6050 would therefore be better, while if one wants to go to MeV energies, S13361-6025 becomes important to avoid such non-linear effects. It should of course be noted that while switching to smaller microcells will allow for a linear response up to higher energies, this does come with the cost of a lower PDE, therefore increasing the lower energy threshold.

It should further be noted that during these tests the shaping time was kept to the minimal value of 12.5~ns while using the peak detection from the Citiroc 1A. This shaping time was previously found to produce the optimum readout for plastic scintillators; by keeping the shaping time the same, it was confirmed that the system can be used to read out mixed scintillator arrays consisting of both GAGG and plastic scintillators, thereby allowing it to be used for example as a small $\gamma$-ray polarimeter. Further tests are needed to see if the performance for GAGG improves when increasing the shaping time.

As the FEE system was confirmed to work well for a scintillator like GAGG which has a decay time of $\approx50\,\mathrm{ns}$ the system will also work well with other fast high-Z scintillators such as pure CsI, LaBr$_3$, and CeBr$_3$. For slower scintillators, such as BGO, additional tests would need to be performed. 

\section{Space Qualification}\label{sec:space_qual}

The SiPM readout has undergone a range of space qualification tests during its development for the POLAR-2 mission. In the following subsections we will describe the thermal vacuum tests, radiation tests, vibration and shock tests separately. All of these tests were passed without any issues.

\subsection{Thermal Vacuum Testing}\label{subsec:tvt}

An assembly of 4 front-end electronics boards coupled to plastic scintillator targets were placed in a vacuum of $\sim 10^{-8}$~mbar for a week in order to test the functionality of the electronics' design. In addition to the 4 front-ends, an additional 5 dummy models were placed with an equal power consumption, thereby simulating the environment for a total of 9 functional front-ends. During these tests, the temperature in the chamber was cycled between $-40^\circ$ C and $+40^\circ$ C with a period of 1600 seconds (800 seconds at each temperature). The tests were performed using the dedicated thermal vacuum test facility at the Max-Planck Institute for Extraterrestrial Physics (MPE) in Garching, Germany.

\begin{figure}[h!]
\centering
 \includegraphics[width=7 cm]{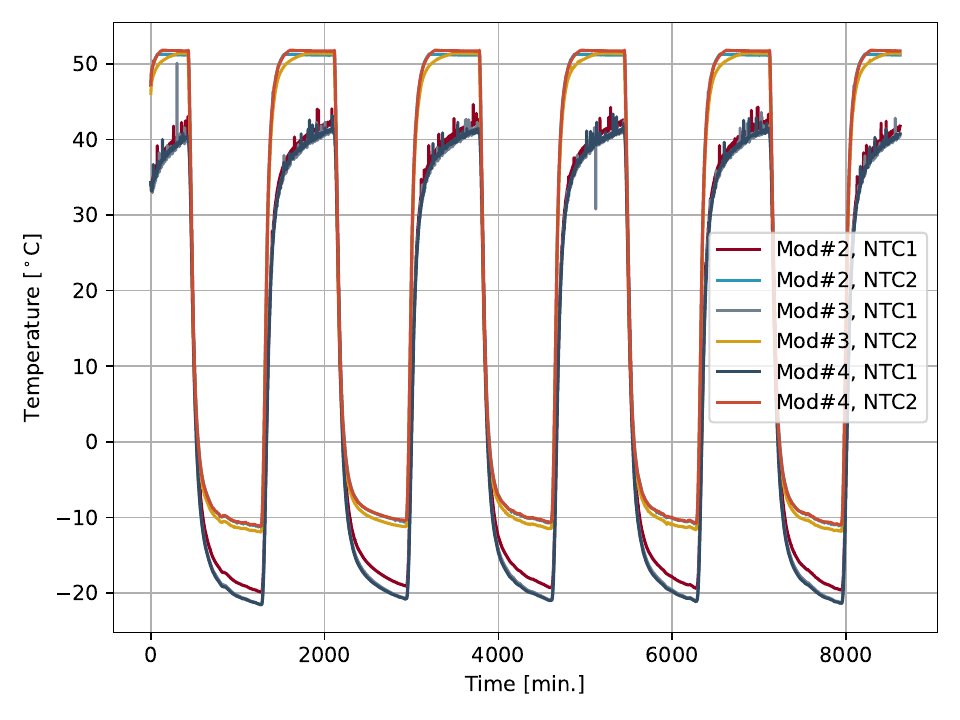}\includegraphics[width=7 cm]{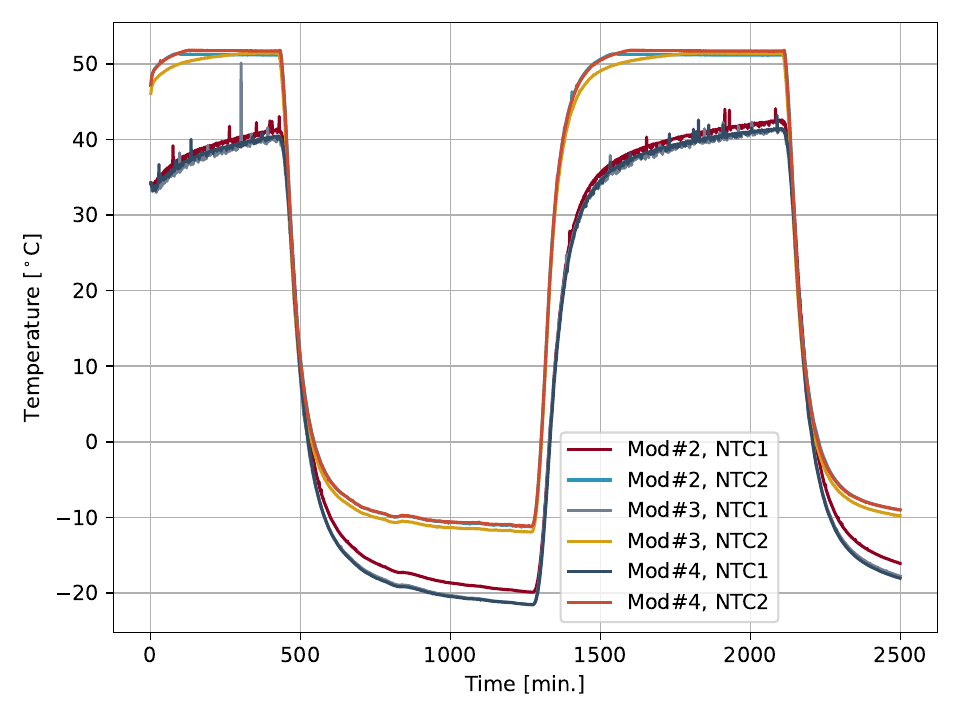}
 \caption{The temperatures as measured on the FPGA (NTC1) and on the SiPM PCB (NTC2) for 3 readout systems as a function of time during a TVT test.}
 \label{fig:TVT_temperature_cycling}
\end{figure}

The housekeeping data, including the temperature monitored by two NTCs on each board, were read out continuously using the four boards simultaneously. One of these NTCs is placed next to the FPGA, while the second one is on the SiPMs array. The temperatures as a function of time are shown for 3 of the front-ends in figure \ref{fig:TVT_temperature_cycling}. It should be noted that one of the front-ends was damaged during the tests as a result of accidentally applying 28 V instead of 3.8 to its input. As a result, the temperature of only 3 boards is shown here. The 3 systems underwent the thermal tests without any issues, thereby passing the test successfully. Based on this test it was decided to replace the initially used power connector to the TigerEye type as described earlier, due to the presence of PVC in the cables originally used with this connector which could cause problems with outgassing.

\subsection{Irradiation}\label{subsec:fee_irr}

One of the main concerns when using COTS components in the design is their resilience to radiation. From all the components used, only the Citiroc 1A ASICs are at TRL 8 \citep{citiroc_1}, thereby indicating that they are radiation tolerant. Neither the IGLOO FPGA, the DC/DC converter used to power the SiPMs, or the ADCs are known to be radiation tolerant. Of particular concern was the DC/DC converter, as initially the C11204-01 by Hamamatsu was considered for use. This DC/DC converter, which is specifically designed for powering SiPMs, failed a basic proton irradiation test resulting in the change to the T3482 DC/DC in the current design. 

\begin{figure}[h!]
\centering
 \includegraphics[height=.34\textwidth]{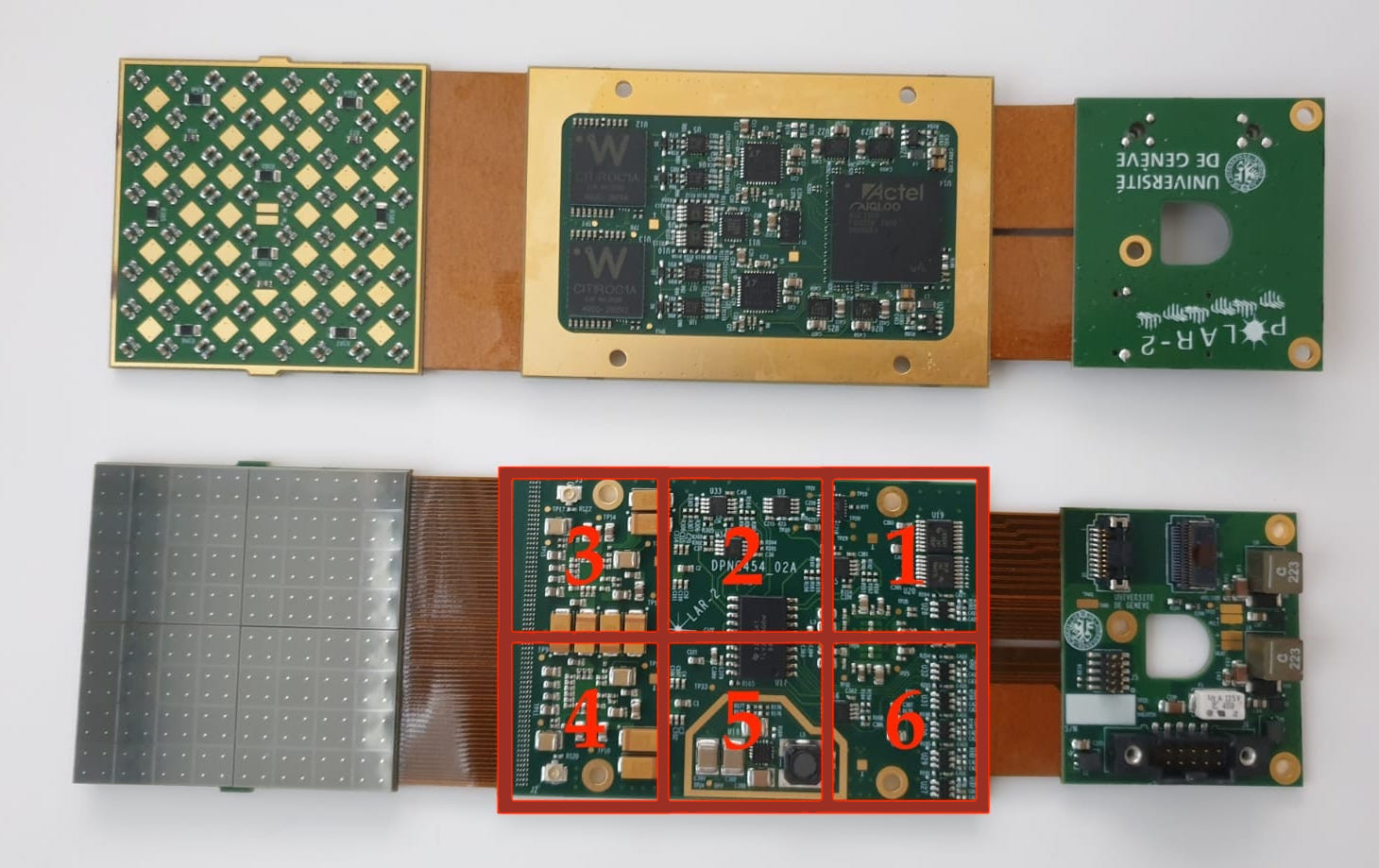}
 \caption{Top and bottom view of the electronics showing the 6 irradiated areas.}
 \label{fig:FEE_irradiation}
\end{figure}

To verify the radiation tolerance of the full design, it was subjected to wide beam proton irradiation. During these tests it was irradiated at the Instytut Fizyki Jadrowej (IFJ)in Krakow, Poland, using a 58 MeV proton beam. Using a brass collimator, a 40 mm diameter beam was shaped into a 25 mm square beam. This allowed to irradiate the central rigid part of the readout in 6 steps. This is indicated in figure \ref{fig:FEE_irradiation} which shows the 6 areas which were irradiated separately to cover the full sensitive part of the readout. Through this irradiation in parts the system was subjected to 3 doses of 0.17 Gy and a fourth dose of 0.25 Gy. Based on simulations of the POLAR-2 detector, this equates to approximately 11 years in low Earth orbit. Data acquisition runs were performed in between each radiation step and no issues or changes in performance, or power consumption, were seen. 

These 11 years in space are calculated based on simulations of POLAR-2, and therefore cannot be directly translated to other missions. Within POLAR-2, the electronics are shielded by several mm of aluminium. However, these levels do indicate that all component on the readout are radiation tolerant to levels equating to several years in low Earth orbit. A future publication presenting more details on these irradiation tests along with an extrapolation of the performance of the readout within POLAR-2 instrument is currently in preparation.

It should of course be noted that the SiPMs are known to be prone to radiation damage. For the POLAR-2 mission these were therefore subject to a detailed dedicated set of tests which are presented in \cite{DeAngelis_2023}.

\subsection{Vibration and Shock Testing}\label{subsec:sin_vibr}

\begin{figure}[h!]
\centering
 \includegraphics[width=.4\textwidth]{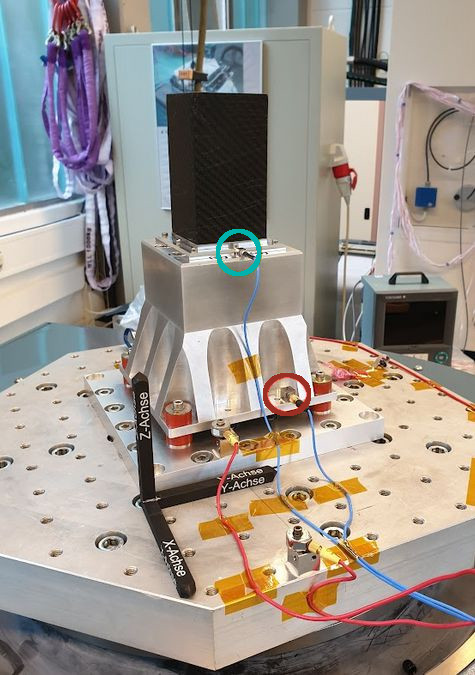}
 \caption{Definition of the X-Y-Z frame with respect to system. The Z axis is defined as the vertical direction (along the scintillator length), and the X and Y directions are defined in the horizontal plane along the scintillators. The readout system is placed in the aluminium frame. Accelerometers are placed on the flange of the module (in green) and near the dampers (in red) in order to monitor the resonance spectrum during and between the different tests \citefig{NDA_thesis}.}
 \label{fig:vibr_setup}
\end{figure}

The system underwent both random and sinusoidal vibration tests on all 3 axes while integrated within a POLAR-2 detector prototype. The tests were performed using the dedicated vibration and shock facility at MPE in Garching, Germany. For this a single functional readout system was used which was tested for functionality both before and after the tests. The board was integrated as indicated earlier in figure \ref{fig:thermomechal_design} where it is coupled with scintillators at the top, while the board is mounted on an aluminium frame. The combination of the readout, the frame itself and the scintillators are mounted within an outer aluminium frame coupled to the vibration and shock table with rubber dampers. The outer frame and dampers are a scaled version of what would be implemented for POLAR-2. The system as mounted on the shaker is shown in figure \ref{fig:vibr_setup}.

The parameters used for the sinusoidal vibrations, applied to all 3 axes, are summarized in table \ref{table:sinusoidal_vibration_specs}. Similarly, those for the random vibrations are summarized in table \ref{table:random_vibration_specs}. Finally, those used for the shocks are shown in table \ref{table:shock_specs}. For the shock tests, the levels were incremented in small steps until the desired levels were measured using sensors placed on the setup. The vibration and shock levels applied here were derived from those applied to the POLAR mission during its final space qualification tests. As the POLAR detector was launched while mounted on the Tiangong-2 spacelab, these levels are relatively high compared to a normal space launch \citep{PRODUIT2018259}.

The system was tested for functionality in between each set of tests, no issues or changes in behaviour were found, thereby indicating that the full system can survive the significant levels of vibration and shock which it will be subjected to during launch. It should be noted that these tests should be seen more as a test of the full POLAR-2 prototype, rather than just the FEE electronics. The mechanics of the prototype provide significant levels of damping to the electronics. Although this level of damping is a result of the mechanics specific to POLAR-2, the results do indicate that the readout can survive such levels when mechanically integrated in a well-designed detector.

\begin{table}[h!]
\begin{center}
\begin{tabular}{|c|c|} 
\hline
Frequency & Qualification \\
range [Hz] & requirement \\ \hline\hline
4-12 & 15~mm \\ \hline
12-17 & 8.8~g \\ \hline
17-75 & 14.5~g \\ \hline
75-100 & 11~g \\ \hline
\end{tabular}
\end{center}
\caption{Requirements used for the sinusoidal vibrations qualification test performed on the 3 axis (x, y, and z), with an acceleration rate of 2~octave/minute.}\label{table:sinusoidal_vibration_specs}
\end{table}




\begin{table}[h!]
\begin{center}
\begin{tabular}{|l|c|c|c|} 
\hline
& \multicolumn{3}{|c|}{Frequency range [Hz]}\\
& 12-250 & 250-800 & 800-2000 \\ \hline\hline
Power spectral density & 6~dB/octave & 0.14~g$^2$/Hz & -9~dB/octave \\ \hline
Total RMS acceleration & \multicolumn{3}{|c|}{11.65~grms} \\ \hline
Test duration & \multicolumn{3}{|c|}{180~s} \\ \hline
Acceleration directions & \multicolumn{3}{|c|}{3 axis} \\ \hline
\end{tabular}
\end{center}
\caption{Requirements used for the random vibrations qualification test.}\label{table:random_vibration_specs}
\end{table}

\begin{table}[h!]
\begin{center}
\begin{tabular}{|l|c|c|} 
\hline
& \multicolumn{2}{|c|}{Frequency range [Hz]}\\
& 100-500 & 500-3000 \\ \hline\hline
Shock response spectral acceleration & 9~dB/octave & 800~g \\ \hline
Test duration & \multicolumn{2}{|c|}{3 times per axis} \\ \hline
Acceleration directions & \multicolumn{2}{|c|}{3 axis} \\ \hline
\end{tabular}
\end{center}
\caption{Requirements used for the shock qualification test.}\label{table:shock_specs}
\end{table}

\section{Conclusion}

A readout system for 64 SiPM channels was developed. The system contains mostly COTS components, allowing to have a cost of several kUSD (approximately 3000~USD when developing 10 units), while also avoiding many of the common export problems found with such electronic systems for space missions. The system furthermore provides a large autonomous readout for 64 SiPM channels with a total power consumption of approximately 1.8~W. 

The system has not shown any significant levels of electronic noise or crosstalk. It was shown to work well both with fast (plastic) and slow scintillators. Despite the relatively low levels of light output of plastic scintillators a charge threshold below 10 keV was possible; using a time threshold this can be significantly lowered. When combining the system with GAGG scintillators, threshold levels of 3.75 keV were achieved without significant system optimization. The higher end of the dynamic range for this combination was furthermore shown to exceed 600 keV, while this can, if needed, be increased to MeV levels. Finally, the system was shown to be capable of accommodating readout rates of several kHz, making it capable of measuring bright GRBs without significant levels of dead time. 

As such, the system can be used both for detectors like POLAR-2 which use only plastic scintillators, as well as spectrometers based on high-Z scintillators. Also, the system can be used to read out combinations of these two scintillators, which can be of use for future polarimeters like SPHiNX \citep{SPHiNX}, or the upcoming evolution of LEAP proposed by some of the authors as an upcoming NASA mission.

The system has furthermore been successfully subjected to space qualification tests, thereby indicating it is capable of surviving both launch conditions as well as long-duration operation in orbit.

Although the system has now been tested for a range of different applications, further usages and applications will be tested in the future. For example, the system remains to be tested with slower high-Z scintillators such as BGO. Furthermore, the system has only been tested with Hamamatsu SiPMs. In the coming year, the system will be tested using various SiPMs from FBK. This includes both cryogenic SiPMs which would be read out through coax cables coupled to the SiPM PCB. This will allow the system to be used for reading out SiPMs in cryostats operated at few Kelvin temperatures. 

Furthermore, the system will be tested using linearly graded SiPMs \citep{LG-SiPM}. Such SiPMs have several (4 or 6) anodes for each channel, where the relative output of the various anodes provides information on the position of the fired microcell. As such, LG-SiPMs can be used to produce large-scale position sensitive detectors, or single LG-SiPM channels can be used to read out many small scintillators such as fibers. The system described here would be capable of reading out up to 16 LG-SiPMs (of the 4 anode readout type), thereby allowing to read out potentially hundreds of scintillating fibers using only 1.8 W.

As the system is already suitable for multi-purpose detectors, its usage is encouraged within the astrophysics community to save development costs for small-scale space missions. In addition, modifications of the system are possible, and encouraged. For example, updates which make use of the Radioroc 2 ASIC should be relatively straightforward.

\section*{Acknowledgements}
We gratefully acknowledge Kurt Dittrich from MPE for his great assistance with the TVT, vibration and shock tests, as well as Lisa Ferro from the University of Ferrara for operating the beam and assisting during the LARIX-A calibration campaign. Additionally, we are very grateful to Coralie Husi and Gabriel Pelleriti from DPNC for their help and expertise respectively on the mechanical and electronics assembly. This work is supported by the AHEAD-2020 Project grant agreement 871158 of the European Union’s Horizon 2020 Program which made the calibration with the GAGG crystal possible. M.K. and N.D.A. acknowledge the support of the Swiss National Science Foundation which supported the initial design and testing phase of the readout system. We gratefully acknowledge the Swiss Space Office of the State Secretariat for Education, Research and Innovation (ESA PRODEX Programme) which supported the development and production. This project has received funding from the European Union's Horizon Europe Research and Innovation programme under Grant Agreement No 101057511 (EURO-LABS) for the irradiation of the FEE.

\appendix

\bibliographystyle{elsarticle-num} 
\bibliography{example}






\end{document}